\documentclass[12pt]{article}
\usepackage{latexsym}
\usepackage{epsfig}
\usepackage{graphicx}
\usepackage{epstopdf}

\usepackage{epsfig,amssymb,amsmath,amscd}

\newcommand{\bq}{\begin{equation}}
\newcommand{\eq}{\end{equation}}
\newcommand{\bqa}{\begin{eqnarray}}
\newcommand{\eqa}{\end{eqnarray}}
\newcommand{\ben}{\begin{enumerate}}
\newcommand{\een}{\end{enumerate}}
\newcommand{\bc}{\begin{center}}
\newcommand{\ec}{\end{center}}
\newcommand{\bqb}{\begin{eqnarray*}}
\newcommand{\eqb}{\end{eqnarray*}}

\newcommand{\qsl}{\rlap / q}

\newcommand{\eesl}{\rlap / \epsilon}
\newcommand{\eeslp}{\rlap / \epsilon'}
\newcommand{\lpsl}{\rlap / l'}
\newcommand{\lsl}{\rlap / l}

\def\pr#1#2#3{Phys. Rev. ${\bf{#1}}$, #2 (#3)}
\def\prl#1#2#3{Phys. Rev. Lett. ${\bf{#1}}$, #2 (#3)}
\def\pl#1#2#3{Phys. Lett. ${\bf{#1}}$, #2 (#3)}
\def\prep#1#2#3{Phys. Rept. ${\bf{#1}}$, #2 (#3)}

\def\np#1#2#3{Nucl. Phys. ${\bf{#1}}$, #2 (#3)}

\def\zp#1#2#3{Z. f. Phys. ${\bf{#1}}$, #2 (#3)}

\def\aop#1#2#3{Annals of Phys. ${\bf{#1}}$, #2 (#3)}

\def\polon#1#2#3{Acta Phys. Polon. ${\bf{#1}}$, #2 (#3)}

\def\jmp#1#2#3{J. Mod. Phys. ${\bf{#1}}$, #2 (#3)}

\begin{document}
\pagenumbering{arabic}
\thispagestyle{empty}
\def\thefootnote{\fnsymbol{footnote}}
\setcounter{footnote}{1}

\begin{flushright}
Sept 27, 2016\\
corrected 1606.08597\\
 \end{flushright}

\vspace{2cm}

\begin{center}
{\Large {\bf Higgs boson form factor effects in $t\bar t$ production by
$W^-W^+$ and $ZZ$ fusion}}.\\
 \vspace{1cm}
{\large G.J. Gounaris$^a$ and F.M. Renard$^b$}\\
\vspace{0.2cm}
$^a$Department of Theoretical Physics, Aristotle
University of Thessaloniki,\\
Gr-54124, Thessaloniki, Greece.\\
\vspace{0.2cm}
$^b$Laboratoire Univers et Particules de Montpellier,
UMR 5299\\
Universit\'{e} Montpellier II, Place Eug\`{e}ne Bataillon CC072\\
 F-34095 Montpellier Cedex 5,  France.\\
\end{center}

\vspace*{1.cm}
\begin{center}
{\bf Abstract}
\end{center}

We study the fusion processes $W^-W^+\to t\bar t$ and $ZZ\to t\bar t$
observable at a future $e^-e^+$ collider and we discuss their sensitivity to an
$Htt$ form factor which may be due to compositeness, in particular when
the $H$ and the top quark have common constituents. We make an amplitude analysis
and illustrate which helicity amplitudes and cross sections for specific final $t\bar t$
polarizations are especially sensitive to this form factor.

\vspace{0.5cm}
PACS numbers: 12.15.-y, 12.60.-i, 14.80.-j

\def\thefootnote{\arabic{footnote}}
\setcounter{footnote}{0}
\clearpage

\section{Introduction}

The peculiar structure of the standard fermion flavours and of the
gauge and Higgs bosons has motivated the search for several types of
explanations \cite{BSMth}. Among them one has the possibility of compositeness
of particles up to now considered as elementary \cite{comp}. The particularly
heavy mass of the top quark has suggested that the correspondingly
large Higgs-top coupling may be due to Higgs and (possibly partial)
top compositeness \cite{Hcomp, Portal, Htop}.

If the Higgs boson and the top quark are (even partially) composite,
they should both have  a form factor.
If in addition their constituents are common  and  do not interact strongly
with a vector boson $V~ ({\rm  with ~}V=\gamma, Z, W)$,
the top quark form factor would not be observed in the usual
$Vtt$ coupling. In such a case only the $Htt$ form factor could be the right place
for the observation of these structures.

But how can we  reach the $Htt$ form factor?
Simply through a process where either one top or the H is far off shell,
with a variable $q^2$. This is a prospect which differs from the search of
anomalous $H$ couplings with on-shell $H$.

Which process would be suitable for the observation of such an $Htt$ $q^2$
dependence?
A final $Ht\bar t$ state (like in $e^-e^+ \to Ht\bar t$) would not be  a priori adequate,
the final $H$ being on shell.
One should for example look for a process with an intermediate H with a variable $q^2$
and a final $t\bar t$ pair.

At a muon collider, the process $\mu^-\mu^+ \to H \to t \bar t$ would be the simplest
one for this purpose, but it is too weak as compared to $\mu^-\mu^+ \to V \to t \bar t$,
which completely dominates the $t \bar t$ production data.
At LHC $gg \to H \to t \bar t$ is also too weak as compared to other (pure QCD)
$gg\to t \bar t$ contributions.

We have therefore  considered the fusion processes $W^-W^+,ZZ \to t \bar t$ which contain
the $W^-W^+,ZZ \to H \to t \bar t$ contributions.
In the Standard Model (SM), these fusion processes have been studied for a long time
\cite{fusion1, fusion2,fusion3}.
But now we want to see the influence of a modification of the Higgs boson
exchange term, in particular through a form factor.
We have therefore recomputed  the $W^-W^+,ZZ \to t \bar t$ amplitudes including, in addition to
the $(W^-W^+,ZZ) \to H \to t \bar t$ contribution, the  corresponding
$b,t$ exchanges in $t$ or $u$ channels,
and the photon and $Z$ exchanges in the $s$ channel. In SM they are all of comparable size.

We have studied the sensitivity of these fusion processes
to the $H$ contribution, especially when there is a $q^2$-dependent
departure from the pointlike SM $Htt$ coupling. In fact this applies
also to a more general situation in which the products $g_{HWW}g_{Htt}$
and $g_{HZZ}g_{Htt}$ get a $q^2$-dependence.

Using these, a peculiar sensitivity appears when the initial vector bosons $W^-W^+$
or $ZZ$ are both longitudinal (LL),
because of the famous gauge cancellations at high energies, which
are now perturbed by Higgs boson form factors.
In order to appreciate the importance of this effect, we illustrate the
behavior of the various amplitudes
and  cross sections for transverse or longitudinal gauge bosons and
for polarized or unpolarized final $t \bar t$ states. We insist on the
importance of separating the various  final state polarizations,  in order to identify
the origin of a possible departure from the SM prediction.

In these illustrations we compare the SM case with pointlike H couplings,
to some examples of $q^2$-dependence. In order to be quantitatively significative
of a form factor effect (and not simply of a change of normalization due
to an anomalous coupling) these examples are constrained to coincide at $q^2=m^2_H$
with the SM value. Moreover,  they should also satisfy
the high energy cancellation of LL amplitudes. We use effective forms satisfying
these two requirements.

Such fusion processes can be observed at a future \cite{Djouadi} $e^-e^+$ collider
in $e^-e^+ \to \nu\bar\nu t \bar t$ and $e^-e^+ \to e^+e^- t \bar t$.
The corresponding cross sections can be easily obtained by integrating the
subprocesses  with the luminosity factors given
by the leading effective weak boson approximation (LEWA).
In the illustrations we use the luminosity factors of \cite{fusion2} and \cite{fusion4}
(where the possibility of polarized $e^-e^+ $ beams has also been considered, which though
does not seem to help for increasing the sensitivity to the $Htt$ form factor).

In the $e^-e^+ \to e^-e^+ t \bar t$ case we add to the $ZZ \to t \bar t$ subprocess,
the background contributions of the $\gamma\gamma\to t\bar t$, $\gamma Z\to t\bar t$,
$Z\gamma\to t\bar t$ subprocesses, which do not contain $H$ effects.

These fusion processes also occur at LHC but with large backgrounds such
that special studies should be done with careful detection characteristics.\\

In Sec. II we treat the $W^-W^+ \to t \bar t$ and $ZZ \to t \bar t$ processes
in SM.
In Sec. III we describe possible expressions for the $Htt$ form factor.
In Sec. IV we discuss and illustrate the behavior of the various amplitudes
and cross sections and their sensitivity to the $Htt$ form factor.
In Sec. V we summarize our comments and list possible improvements and developments.
An Appendix collects the explicit expressions of the various helicity amplitudes.\\

\section{SM analysis of $W^-W^+ \to t \bar t$ and $ZZ\to t \bar t$.}

We express the amplitudes
\bq
W^-(\epsilon,\lambda,l)+W^+(\epsilon',\lambda',l') \to t(\tau,p)+\bar t(\tau',p') ~~,
\label{WW-amp}
\eq
in terms of the $W^{\mp}$ polarization vectors, helicities and momenta respectively denoted as
$(\epsilon,\lambda,l)$ and $(\epsilon',\lambda',l')$;
as well as the $t,\bar t$ helicities and momenta denoted as $(\tau,p)$ and $(\tau',p')$
(see the Appendix).  By $\theta$ we denote the center of mass  angle between the
$W^-(\epsilon,\lambda,l)$ and  $t(\tau,p)$ directions of motion, and we also define
\bqa
&& s=q^2=(l+l')^2=(p+p')^2 ~~,~~ \nonumber \\
&& t=q_t^2=(l-p)^2=(l'-p')^2  ~~,~~  u=q_u^2=(l'-p)^2=(l-p')^2 ~~. \label{stu-var}
\eqa
The Born level amplitude is written as
\bq
A=A^{WW}_u+A^{WW}_{s,\gamma}+A^{WW}_{s,Z}+A^{WW}_{s,H} ~~, \label{Born-WWamp}
\eq
in terms of the four diagrammatic contributions induced by

\noindent
the \underline{u-channel bottom exchange}
\bq
A^{WW}_u={-e^2g^L_{Wtb}\over q^2_u-m^2_b}\bar u(t)[\eeslp P_L(\qsl_u+m_b)\eesl P_L]v(\bar t)
~~,\label{Born-WW-u-bottom}
\eq
with $g^L_{Wtb}=1/( s_W\sqrt{2})$  ~~, \\

\noindent
the \underline{s-channel $V=\gamma,Z$ exchange}
\bq
A^{WW}_{s,V}={-e^2g_V\over s-m^2_V+im_V\Gamma_V}\bar u(t)[2(\epsilon.l')\eeslp+2(\epsilon.l)\eesl-
(\epsilon.\epsilon')(\lpsl-\lsl)](g^L_{Vt}P_L+g^R_{Vt}P_R)v(\bar t) ~~,\label{WWV}
\eq
with
\bq
g_{\gamma}=1 ~,~ g_Z={c_W\over s_W} ~,~ g^L_{\gamma t}=g^R_{\gamma t}={2\over3}~,~
g^L_{Z t}=~{1\over2s_Wc_W}\left (1-{4s^3_W\over3} \right )~,~ g^R_{Z t}=-{2s_W\over3c_W} ~~,
\label{Ztt}
\eq \\

\noindent
and \underline{s-channel H exchange}
\bq
A^{WW}_{s,H}={-e^2g_{HWW}g_{Htt}\over s-m^2_H+im_H\Gamma_H}(\epsilon.\epsilon')[\bar u(t)v(\bar t)]
~~, \label{WWH}
\eq
with
\[
g_{Htt}=-{m_t\over2s_Wm_W}~~, ~~g_{HWW}={m_W\over s_W} ~~.
\]\\

The 36 helicity amplitudes $F_{\lambda,\lambda',\tau,\tau'}(s,\theta)$ implied by (\ref{Born-WWamp}),
are listed in the Appendix using the standard forms of
Dirac spinors and vector boson helicities\footnote{The transverse (T) $\lambda,\lambda'=-1,+1$ and
longitudinal (L) $\lambda,\lambda'=0$ helicities are obtained from the corresponding
$W^{\mp}$ polarization vectors $\epsilon,\epsilon'$.} \cite{JW}.
These contain  the 10 helicity conserving (HC) amplitudes which satisfy
the  rule \cite{hc} $\lambda+\lambda'=\tau+\tau'$, and the 26 helicity violating (HV)
ones which violate it. Only  6 of them  contain an $H$ exchange
contribution. Note that CP conservation imposes  the relation
\bq
F_{\lambda,\lambda',\tau,\tau'}(s,\theta)=F_{-\lambda',-\lambda,-\tau',-\tau}(s,\theta) ~~.
\label{CP-constraint}
\eq

In terms of the aforementioned helicity amplitudes,
the cross sections for specific $W^-,W^+$ and $t,\bar t$ helicities are written as
\bq
{d\sigma (W^-_\lambda W^+_{\lambda'}\to t_\tau \bar t_{\tau'})\over d\cos\theta}
={3p_t\over 32\pi l_W s}
|F_{\lambda,\lambda',\tau,\tau'}(s,\theta)|^2  ~~, \label{dsigmaWW-pol}
\eq
involving (\ref{stu-var}) and the magnitudes of the $t$ and $W$ c.m. momenta
\bq
p_t=\sqrt{{s\over4}-m^2_t}~~, ~~l_W=\sqrt{{s\over4}-m^2_{W}} ~~. \label{t-W-3momenta}
\eq

We next turn to the corresponding cross sections for  $e^-e^+\to \nu\bar\nu t\bar t$  obtained by
multiplying each ($\lambda,\lambda'$) contribution to (\ref{dsigmaWW-pol})  by the corresponding
luminosity factor
\bq
L_{\lambda, \lambda'}(y)=\int_{y}^1 {dx \over x}f_{W^- , \lambda}(x)f_{W^+, \lambda'}
\left ({y\over x} \right ) ~~, \label{lumi}
\eq
computed with the $W^{\mp}$ distribution functions as given in \cite{fusion2,fusion4}, with
$y=s/(4E^2)$ and $E$ being the c.m. $e^\mp$ beam energy. The integration limits
($y,1$) can be modified to ($x_{min},x_{max}$) when kinematical detection cuts
are applied. As in the parton model, in the leading approximation the $W^{\mp}$
 vector bosons are emitted along the $(e^-,~e^+)$ axis,
 so that the final $(t,~\bar t)$ direction in their center of mass
is also given by the same angle $\theta$, as the one defined just before (\ref{stu-var}).

The $e^-e^+$ cross sections
\bq
{d\sigma_{e^-e^+} (\tau,\tau')\over dyd\cos\theta}=\sum_{\lambda,\lambda'}L_{\lambda,\lambda'}(y)
{d\sigma(W^-_\lambda W^+_{\lambda'}\to t_\tau \bar t_{\tau'})\over d\cos\theta} ~,
\label{dsigmaWW-pol-ee}
\eq
are also computed for specific final $t,\bar t$ polarizations or, by summing them,
for an unpolarized final state.\\

 For the \underline{$ZZ\to t \bar t$ process}, the notation is the same as for $W^-W^+$.
Explicitly, we now have
\bq
Z(\epsilon,\lambda,l)+Z(\epsilon',\lambda',l')\to t(\tau,p)+\bar t(\tau',p') ~~,
\label{ZZ-amp}
\eq
but different diagrams now occur involving no s-channel $(\gamma,Z)$ exchange, but
both $u$- and $t$-channel top exchange contributions given by
\bqa
A^{ZZ}_u &=& {-e^2\over q^2_u-m^2_t}\bar u(t)[\eeslp (g^L_{Zt}P_L+g^R_{Zt}P_R)
(\qsl_u+m_t)\eesl (g^L_{Zt}P_L+g^R_{Zt}P_R)]v(\bar t) ~~, \nonumber \\
A^{ZZ}_t &=&{-e^2\over q^2_t-m^2_t}\bar u(t)[\eesl (g^L_{Zt}P_L+g^R_{Zt}P_R)
(\qsl_t+m_t)\eeslp (g^L_{Zt}P_L+g^R_{Zt}P_R)]v(\bar t) ~~, \label{Born-ZZ-u-t}
\eqa
\noindent
as well as  $s$-channel $H$ exchange contributions
\bq
A^{ZZ}_{s,H}={-e^2g_{HZZ}g_{Htt}\over s-m^2_H+im_H\Gamma_H}(\epsilon.\epsilon')[\bar u(t)v(\bar t)]
\label{ZZH} ~~,
\eq
involving $g_{Htt}=-m_t/(2s_Wm_W)~~,~~g_{HZZ}=m_Z/(s_Wc_W)$.

The above $u-$ and $t$-channel top exchange expressions are also used for
describing the $\gamma\gamma\to t\bar t$, $\gamma Z\to t\bar t$, $Z\gamma\to t\bar t$
processes, by  adapting the $g^{L,R}_{Zt}$ couplings in (\ref{Ztt}).
These processes do not involve the $H$
exchange diagram and the $Htt$ form factor. They constitute a background
to be added to $ZZ\to t\bar t$, when we consider the $e^+e^-\to e^+e^- t\bar t$ process.

The corresponding $e^-e^+\to e^-e^+ t\bar t$ cross section is obtained
as in the previous $W^-W^+$ case and (\ref{lumi}) and (\ref{dsigmaWW-pol-ee}),
by using the $Z$ and $\gamma$ distribution functions
also given in \cite{fusion2,fusion4}. In the present work we
give  results with and without the $\gamma\gamma\to t\bar t$, $\gamma Z\to t\bar t$,
$Z\gamma\to t\bar t$ backgrounds.\\

\section {Higgs boson form factor effects}

The simplest Higgs boson form factor effect comes through the products
of couplings $g_{HWW}g_{Htt}$, $g_{HZZ}g_{Htt}$ in (\ref{WWH}) and (\ref{ZZH}),
when they get an anomalous $q^2$ dependence. This may simply be due
to the presence of an $Htt$ form factor in $g_{Htt}$, or to the simultaneous presence
of $Htt$ and of $HWW$, $HZZ$ form factors.

Let us first insist on the difference between such a form factor effect
and the presence of anomalous couplings. Anomalous couplings are usually described
by effective high dimension operators generated by new
interactions with a new physics scale $\Lambda$ higher than the available
energy. In the case of $Htt$ see for example
\cite{Saavedra}. The effect of each of such operators is a departure from the SM couplings
already appearing on shell and behaving like powers of $q^2/\Lambda^2$.

Our form factor description of compositeness effects assumes that the on-shell SM
value is reproduced by the compositeness structure, but that a complete $q^2$
dependence is generated even within the new physics domain.
It could possibly be described by an involved sum of effective operators with different dimensions.

As we only study the observability of effects of the presence of a form factor,
and not of a single anomalous coupling
modifying the SM on-shell value of the coupling, we  use an effective form factor
generating some $q^2$ dependence, but imposing the SM value at $q^2=m^2_H$.

In the spirit of compositeness, a typical  $q^2$ dependence of the $Htt$ vertex
may be given by an $STS$ or a $TST$ triangular loop
(where $S$ and $T$ are new scalar or fermion constituents of mass $m$).

The crude result, imposing the SM normalization value at $q^2=m^2_H$ to
the product of couplings, would be
\bq
g^0_{Htt}(q^2)=g^{SM}_{Htt}{C(q^2)\over  C(m^2_H)} ~~, \label{crude}
\eq
where in the STS case we have
\bq
C(q^2)=m_t(C_0(q^2)+C_{12}(q^2)-C_{11}(q^2))+mC_0(q^2) ~~, \label{STS}
\eq
in terms of Passarino-Veltman functions \cite{PV}, while in the TST case we have
\bqa
C(q^2) &= & \kappa(q^2)+q^2C_{12}(q^2)+m^2C_0(q^2)+2m_tm(C_{12}(q^2)-C_{11}(q^2)) ~~, \nonumber \\
\kappa(q^2)& = & m^2_t(C_{21}(q^2)+C_{22}(q^2))-2C_{23}(q^2))+q^2C_{23}(q^2)+4C^r_{24}(q^2) ~~,
\label{TST}
\eqa
where $C^r_{24}(q^2)$ is a divergence free quantity (for example as given by the
SRS scheme \cite{SRS}).

However such  choices would destroy the cancellation of LL amplitudes at high
$(s, t, u)$-values, so that they  cannot correspond to a viable
model. There are various ways to recover the cancellation. They may  for example
depend on the assumption of partial or total fermion compositeness, in addition to
Higgs compositeness. This will correspondingly affect the
$b$, $t$ and $V$ exchanges and/or $H$ exchange amplitudes.
In any case these additional contributions should combine with the
contribution involving the Higgs form factor in such a way that the total
satisfies unitarity, i.e. does not explode at high energy.
There may be various such acceptable results
which would quantitatively differ from the standard case. The minimal change would
correspond to an effective value which becomes similar to the SM one at high $q^2$.
As our aim is only to study the sensitivity
of the considered processes to the occurrence of form factors, we choose an arbitrary
effective $q^2$ expression which satisfies the above minimal requirement. Any other
choice should give larger differences with the SM prediction.

One example is
\bq
g^{eff}_{Htt}(q^2)=g^{SM}_{Htt}\left \{{C(q^2)\over C(m^2_H)} +{q^2-m^2_H\over q^2+4m^2}
(1-{C(q^2)\over C(m^2_H)})\right \} ~~, \label{effective}
\eq
where either (\ref{STS}) or (\ref{TST}) are used,
respectively producing the STS and TST  {\it effective} model variations.

Another (trivial) possibility could be the occurrence of one (or more) resonance ($H'$),
located at intermediate $q^2$ values.
The corresponding contribution
\bq
A^{WW,ZZ}_{s,H'}={-e^2g_{H'XX}g_{H'tt}\over q^2-m^2_{H'}+im_{H'}\Gamma_{H'}}
(\epsilon.\epsilon')[\bar u(t)v(\bar t)] ~~, \label{resonance}
\eq
\noindent
with  $X= W ~ {\rm or}~ Z $,  corresponds to a form factor effect written as
\bq
g^{eff}_{Htt}(q^2)=g_{Htt}+g_{H'tt}{g_{H'XX}(q^2-m^2_{H}+im_{H}\Gamma_{H})
\over g_{HXX}(q^2-m^2_{H'}+im_{H'}\Gamma_{H'})} ~~,
\eq
and satisfying  the constraint
\bq
g_{Htt}g_{HXX}+g_{H'tt}g_{H'XX}=g^{SM}_{Htt}g^{SM}_{HXX} ~~,
\eq
so that the high energy cancellation is obeyed.

The shapes of the Re and Im parts of such form factors are shown in
Figs. 1(a) and (b) where we have arbitrarily chosen three examples:  the {\it crude  STS}  case  based
on (\ref{crude}) and (\ref{STS}), the {\it effective} case based
on (\ref{STS}) and (\ref{effective}), both with a constituent mass $m=0.5$ TeV,  and the
 {\it resonance} case based on (\ref{resonance}) and the  mass-choice\footnote{This choice
 was initiated by a signal seen in the  2015 data  \cite{ATLAS750, CMS750},
 which meanwhile has disappeared \cite{no750}. In any case though it may still be used as an example.}
 $m_{H'}=0.75$ TeV. In these figures, one can see in particular the differences of the high energy
behavior  in both real [Fig.1(a)] and imaginary [Fig.1(b)] parts of the form factor.
The corresponding effects in amplitudes and cross sections are illustrated below.\\

\section{Properties of amplitudes and cross sections.}

The basic results for total SM cross sections
have been given in \cite{fusion2}. Here we  present a detailed amplitude analysis,
in order to provide an understanding of the modifications resulting from the presence
of new $q^2$ dependencies in the $H$ exchange term.

\subsection{The $W^-_\lambda W^+_{\lambda '}\to t_\tau \bar t_{\tau '}$ process.}

We discuss separately the pure transverse (TT), the mixed (TL, LT)
and the pure longitudinal (LL) amplitudes.  Since in SM there are specific strong cancellations
between the various terms contributing to the (TL, LT)  and to the
(LL) amplitudes, they  correctly behave at high energy.

During this analysis we also check the helicity conservation rule
valid in SM at tree level \cite{hc}, but not in the presence of  arbitrary form factors.

The ten HC amplitudes are
\bq
F_{-+-+} , F_{-++-} ,  F_{+--+} , F_{+-+-} ,  F_{0---}= F_{+0++} , F_{-0--} = F_{0+++},
F_{00-+}  , F_{00+-} ~~ , \label{WW-HC-SMamp}
\eq
while the 26 HV ones, which are suppressed at high energy in the pure SM case, are
\bqa
&&  F_{---+}=F_{++-+}, F_{--+-}=F_{+++-}, F_{-+--}=F_{-+++}=F_{+---}=F_{+-++} , \nonumber \\
&& F_{0-++}=F_{+0--}, F_{0--+}=F_{+0-+},  F_{0-+-}=F_{+0+-},  F_{-0++}=F_{0+--}, \nonumber \\
&& F_{-0-+}=F_{0+-+} ~,~  F_{-0+-}=F_{0++-}~,   \label{WW-HV1-SMamp} \\
&& F_{----}=F_{++++} ~,~ F_{--++}=F_{++--}~,~ F_{00++}=F_{00--} ~~.  \label{WW-HV2-SMamp}
\eqa

Figure 2   presents the HC amplitudes  (\ref{WW-HC-SMamp}) in the upper panels,
 while  the HV amplitudes of (\ref{WW-HV1-SMamp}) are shown in the lower panels.
 The left panels refer to TT $W^-W^+$, while right panels always involve at least one
 longitudinal $W^\mp$. Both of these sets do not involve any s-channel H exchange,
so there is no effect of the form factor.
At high energy the HC amplitudes
have weak energy dependencies and no or negligible imaginary parts.
The leading ones are  $F_{-+-+}$, $F_{+--+}$, $F_{0---}=F_{+0++}$ and $F_{00-+}$,
with a high energy limiting absolute value of the order
of 0.05 to 0.6.  Concerning particularly $F_{00-+}, F_{00+-}$, we note that they acquire their
high energy  magnitude after a strong  cancellation between the
u-channel bottom exchange and s-channel $\gamma,Z$ exchange contributions.
We also note that the HV amplitudes  in the lower panels of Fig.2, involve  high energy
cancellations    among a  u-channel bottom exchange and an s-channel $\gamma,Z$ exchange,
and decrease with the energy. These HV amplitudes  are smaller than the HC ones shown
in the upper panels.

We next turn to the six HV amplitudes appearing in (\ref{WW-HV2-SMamp}), which
are the only ones  receiving  an $H$ exchange contribution, and being
therefore sensitive to the $Htt$ form factor. These are shown in Fig.3, where
the standard amplitudes involving the SM pointlike
$Htt$ coupling, are compared to those induced by the examples of anomalous $Htt$
form factors shown in Fig.1. These $H$ sensitive amplitudes are of variable size.
The four (TT) ones listed in the upper and middle panels of Fig.3,
 are almost 10 times smaller than the leading "no H" ones shown in Fig.2.

The two (LL) ones, appearing in the lower panels of Fig.3, can reach much larger values.
One can see the specific energy dependencies in the {\it effective} and {\it resonance} cases.
Imaginary parts  may be important above the "new" threshold or around the resonance.
The sensitivity of the  four (TT) amplitudes to the
$Htt$ form factor is not as strong as the sensitivity of the two (LL) ones.
The chosen form factor leads to modifications of these amplitudes
for $s\gtrsim (m^2, ~ m^2_{H'})$  (the new scale) and mostly at higher energies, although
they satisfy the cancellation at very high energies.

In addition to the modification of the amplitudes
around $(m^2, ~ m^2_{H'})$ by the form factor effect,
a strong absence of cancellation effect would appear at high energy
in the {\it crude STS} form factor choice.
With the other choice (the "{\it effective}" one
satisfying the cancellation constraint) there still remains a strong departure
from the SM prediction. So these LL amplitudes, $F_{00++}=F_{00--}$, are the clearest
source of large sensitivity to the $Htt$ form factor.\\

{\bf Form factor sensitivity of the
$d \sigma (e^-e^+ \to \nu\bar\nu t_\tau \bar t_{\tau'})/d\cos\theta$ ~cross sections:}
These cross sections would  reflect the above amplitude
properties. In this case,  for each final $t_\tau\bar t_{\tau'}$ polarization, we have to sum
the different $W^-_\lambda W^+_{\lambda'}$ initial state contributions,
each probability being multiplied
by the corresponding luminosity (\ref{lumi}). This sum involves all HV and HC amplitudes, even
those which do not contain the $H$ contribution,  thereby diminishing the relative size of
the form factor effect. In Fig.4 we thus present the following illustrations:
\begin{itemize}

\item
In the upper panel we show the energy dependencies of the left-right ($\tau=-\tau'=-1/2 $)
 and right-left ($\tau=-\tau'=1/2 $) differential cross sections at $\theta=60^\circ$.
There exists  no H contribution to these quantities.

\item
In the middle panels we show the left-left ($\tau=\tau'=-1/2 $) and right-right $\tau=\tau'=1/2 $)
differential cross sections, again at $\theta=60^\circ$.

\item
Finally, in the lower panels we give the energy (at $\theta=60^\circ$)
and the angular (at $\sqrt{s}=1$TeV)
dependencies, when the $t\bar t$-helicities  are not observed.

\end{itemize}

The results in the middle and lower panels of Fig.4 do depend in the $Ht\bar t$ form factor.
In them, one recognizes the strong threshold or resonance  effects  seen in the amplitudes in Fig.3
around $s\simeq(m^2, m^2_{H'})$ and  higher energies.

As expected, the form factor effects
in the lower panels of Fig.4 are somewhat smaller than those in the middle panels, since 
the first also involve amplitudes insensitive to $H$.

In the right lower panel of Fig.4 we also show the angular dependence in the unpolarized case
at $\sqrt{s}=1$ TeV.
One can see that the $H$ contribution gives an additional typical constant angular dependence,
which differs from the backward peaking of the dominating $u$ channel exchange.
The effect is therefore mainly localized in the forward and central domain.\\

Finally we have also computed the integrated cross sections
for some examples of the $e^{\mp}$ c.m. energy $E_e$,  with the $t\bar t$ invariant mass
chosen larger than some minimal value $\sqrt{s}>\sqrt{s_{min}}$.
In this case we have imposed an angular cut in order to eliminate the main background coming
from photon radiation, in agreement with the study of \cite{Larios}. Thus

\begin{itemize}
\item
for $E_e=1$ TeV and $\sqrt{s_{min}}=2m_t$ or 1 TeV, one gets integrated cross sections
of 0.10 or  0.03 fb in SM, and 6.1 or 0.17 fb in the presence of  the form factor;

\item
while  for $E_e=3$ TeV and $\sqrt{s_{min}}=2m_t$ or 1 TeV,
one gets integrated cross sections
of 3.8 or 1.2 fb in SM,  and 72. or 5.8 fb in the presence of  the form factor.

\end{itemize}

These dependencies can easily be understood from the shapes of the corresponding
energy dependencies of the SM and the form factor effects shown in Fig.4.
Accordingly, for the energies of a future collider,
(see for example Fig.1 of \cite{colliders} and its Refs.[1-4]
a luminosity of $10^{35} {\rm cm^{-2} s^{-1}}$ would give $10^2$ to $10^5$ events
per year (depending on the energies and  the cuts) and a large observability of form factor effects.

\subsection{The  $Z_\lambda Z_{\lambda'} \to t_\tau \bar t_{\tau'}$ process.}

The properties of the various helicity amplitudes
for $Z_\lambda Z_{\lambda'} \to t_\tau \bar t_{\tau'}$ are globally similar
to those of the $W^-W^+$ case.
The cancellations  for longitudinal amplitudes are also
similar to those for  $W^-W^+$, but  it now occurs
between the $t$ and $u$ channel top exchange (there is no $\gamma,Z$ exchange),
and also with the $s$-channel H exchange in the $F_{00++}=F_{00--}$ case.
These later amplitudes are  the leading ones, with the
larger sensitivity to the form factor.

Following a similar a procedure as for the previous case\footnote{Compare Fig.2}, we present in
Fig.5   the insensitive to the H form factor amplitudes, with  upper panels
describing  the tree level SM  HC amplitudes,
\bq
F_{-+-+} , F_{-++-} ,  F_{+--+} , F_{+-+-} ,  F_{0---}= F_{+0++} , F_{-0--} = F_{0+++},
F_{00-+}  ~~ , \label{ZZ-HC-SMamp}
\eq
and the lower panels showing the HV ones
\bqa
&&  F_{---+}=F_{++-+}, F_{--+-}=F_{+++-}, F_{-+--}=F_{-+++}=F_{+---}=F_{+-++} , \nonumber \\
&& F_{0-++}=F_{+0--}, F_{0--+}=F_{+0-+},  F_{0-+-}=F_{+0+-},  F_{-0++}=F_{0+--}, \nonumber \\
&& F_{-0-+}=F_{0+-+} ~,~  F_{-0+-}=F_{0++-}~.   \label{ZZ-HV1-SMamp}
\eqa
The amplitudes in Fig.5 are all real.

Correspondingly, Fig.6 shows the three H-form factor sensitive amplitudes in
\bq
 F_{----}=F_{++++} ~,~ F_{--++}=F_{++--}~,~ F_{00++}=F_{00--} ~~,  \label{ZZ-HV2-SMamp}
\eq
in the upper, middle and lower panels respectively. The same form factor models as in Fig.1 are used.
Left and right panels are respectively giving the real and imaginary parts of these amplitudes.

A source of differences between the $ZZ$
and the $W^-W^+$ case is the weaker $Ztt$ couplings which lead to a weaker SM (non-H)
contribution and therefore a relatively larger sensitivity to the $Htt$ form factor,
both around $s\backsimeq(m^2, ~m^2_{H'})$ and  higher energy.

But the largest differences come from the $ZZ$ symmetry, which
renders the angular distribution symmetric with respect to
$\theta\to \pi-\theta$.
This leads in particular to the presence of
both forward and backward peaking of the non-H contribution.
Both features lead to a large sensitivity to $H$ exchange in the central domain.

A priori the comparison of the $W^-W^+$ and $ZZ$ processes would contribute
to the identification of a possible nonstandard $Htt$ effect.\\

{\bf Form factor sensitivity of the
$d \sigma (e^- e^+ \to e^- e^+  t_\tau \bar t_{\tau'})/d\cos\theta$ ~cross sections:}
The new point is now that the $ZZ$ process gets large background effects from
$\gamma\gamma\to t\bar t$, $\gamma  Z\to t\bar t$ and $Z\gamma\to t\bar t$
processes which have no $H$ contribution.
This background is essentially dominated by the $\gamma\gamma\to t\bar t$
contribution. It can however be reduced by detecting the final $e^-e^+$
and making an angular cut rejecting their big forward contribution.

In the following illustrations we have taken a background example
obtained with a tentative cut at $\theta=0.1$. This reduces the background which is then
only one order of magnitude larger than the SM pure $ZZ$ contribution.
We thus  successively show the effects in the polarized (Figs.7 and 8) and in the unpolarized
(Fig.9) $t\bar t$ cross sections.

Figure 7  presents H-independent cross sections for $\tau=-\tau'$ showing the relative
importance of the background contribution.

Figure 8 presents then the H-dependent cross sections for $\tau=\tau'$ showing that
the modifications of the ZZ contributions due to H form factors can nevertheless be
observable (locally in the case of a resonance or in the high energy behavior in the case
of an effective form) even with the presence of the background.

Finally in Fig.9 we present the unpolarized  cross sections. The upper panels give energy dependencies and the
lower panels the angular ones. It appears that
even in this unpolarized case, effects of effective or resonance contributions can then be clearly seen,
at the (20\%-50\%) level with the chosen parameters,  mostly in the central angular region.\\

As in the previous subsection, we have also computed the integrated cross sections for 
$ZZ+{ \rm background}$ 
with the same examples of electron $e^{\mp}$ energy $E_e$, minimal
$t\bar t$ invariant mass $\sqrt{s_{min}}$ and cuts. Thus

\begin{itemize}
\item
for $E_e=1$ TeV and $\sqrt{s_{min}}=2m_t$ or 1 TeV, one gets total (ZZ+background)
integrated cross sections values
of 0.92 or 0.023 fb in SM, and 0.92 or 0.031 fb with the form factor;

\item
while for $E_e=3$ TeV and $\sqrt{s_{min}}=2m_t$ or 1 TeV, one gets integrated cross sections values
of 7.2 or 0.5 fb in SM, and 7.2 or 0.8 fb with the form factor.
\end{itemize}

These dependencies can also be understood from the shapes of the corresponding
energy dependencies in Fig.9, in particular the large sensitivity to the
minimal $t\bar t$ invariant mass.
Although the sensitivity to the form factor effects is weaker than in the $WW$ case,
these $ZZ+{ \rm background}$ results,
with $10^2$ to $10^4$ events in a future collider \cite{colliders},
could still correspond to observable situations.

Of course these illustrations just correspond to arbitrary examples. Detection characteristics
should be adapted to the real observations.\\

\section{Final comments and  possible  developments.}

We have shown what could be the effect of a $Htt$ form factor on
the energy and angular dependencies of the amplitudes and cross sections
of the $W^-W^+\to t\bar t$ and $ZZ\to t\bar t$ fusion processes, especially
when identifying the final $t\bar t$ polarizations.

For the illustrations we have taken simple expressions of form factors,
for example arising from common constituents of the Higgs boson and
the top quark. We have chosen effective expressions satisfying normalization
constraints and leading to acceptable high energy behavior of the amplitudes.
For comparison we have also shown the spectacular effect generated by a new resonance form.

By using the relevant LEWA functions of \cite{fusion1,fusion2,fusion4} we have seen
how the above sensitivity of the $W^-W^+\to t\bar t$ and $ZZ\to t\bar t$ subprocesses
is transmitted to the $e^-e^+ \to \nu\bar\nu t \bar t$ and $e^-e^+ \to e^-e^+ t \bar t$
cross sections. For the $ZZ$ case we have added the background contributions
of the $\gamma\gamma\to t\bar t$, $\gamma Z\to t\bar t$, $Z\gamma\to t\bar t$ subprocesses
which do not contain $H$ effects. This background can be reduced by applying angular
cuts to the detection of the final $e^-e^+$.

A large sensitivity to the $Htt$ form factor is observed in the energy
and  angular dependencies of these cross sections. With the expected values
of the future colliders energy and luminosity, a large number of events
is expected leading to a good observability of the form factor effects.
Measurements of the final $t\bar t$ polarizations would increase this sensitivity
and help for the identification of the origin of a possible departure
from the SM predictions.

Applications to LHC may also be considered with $W^-W^+\to t\bar t$ and
$ZZ\to t\bar t$ fusion, after emission of $W$ and $Z$  by $q$ and $\bar q$ partons;
but there are many other subprocesses creating
$t\bar t$ pairs which will overwhelm these fusion ones. Involved sets of
detection features may help to separate them, but this is beyond our competence.\\

\vspace*{1.cm}

\renewcommand{\thesection}{}

\section{\bf  Appendix: The helicity amplitudes $F_{\lambda,\lambda',\tau,\tau'}(s,\theta)$}

\setcounter{equation}{0}
\setcounter{subsection}{0}
\renewcommand{\thesubsection}{A.\arabic{subsection}}
\renewcommand{\theequation}{A.\arabic{equation}}

The usual Dirac spinor decomposition is made for $\tau,\tau'$ helicities
of the top and antitop with momenta $p^{\mu}=(p^0_t,p_t\sin\theta,0,p_t\cos\theta)$
and $p^{'\mu}=(p^0_t,-p_t\sin\theta,0,-p_t\cos\theta)$:
\bq
\bar u(p,\tau)=\sqrt{E_t+m_t}(\chi^+_{\tau}~, ~-2\tau r_t\chi^+_{\tau})~,~
v^T(p',\tau')=-\sqrt{E_t+m_t}(r_t\chi_{\tau'}~, ~-2\tau'\chi_{\tau'})~,
\eq
with
\bq
r_t={p_t\over E_t+m_t}~~,~~\chi^+_{+}=\left (\cos{\theta\over2}~,~\sin{\theta\over2} \right )
~,~ \chi^+_{-}=\left (-\cos{\theta\over2}~,~\sin{\theta\over2} \right )~.
\eq
For the No.1 gauge boson according to the standard Jacob-Wick (JE) convention
\cite{JW} $V=W^-,Z$, with momentum $l^{\mu}=(l^0,0,0,l)$, the transverse and longitudinal
polarization vectors are respectively given by
\bq
\epsilon^{\mu}(l,\lambda)=\left (0,{-\lambda\over\sqrt2},{-i\over\sqrt2},0 \right )
~,~ \epsilon^{\mu}(l,0)=\left ({l\over m_V},0,0,{l^0\over m_V} \right) ~,
\eq
while for the No.2 $V'=W^+,Z$, with momentum $l^{\prime \mu}=(l^0,0,0,-l)$, the corresponding
polarization vectors are
\bq
\epsilon^{\prime \mu}(l',\lambda')=\left (0,{\lambda'\over\sqrt2},{-i\over\sqrt2},0 \right )
~,~\epsilon^{\prime \mu}(l',0)=\left ({-l\over m_V},0,0,{l^0\over m_V} \right )~.
\eq

The contributions to the helicity amplitudes are of three kinds 
(see Sec. II).

\begin{itemize}

\item

{\bf  u- and t-channel exchange.}

 For $W^-W^+$ we only have  u-channel bottom exchange leading to
\bq
A^{WW}_u\to -~{e^2g^{2L}_{Wtb}(E_t+m_t)\over 2(u-m^2_b)} \{T^{1L}_u+T^{2L}_u+T^{3L}_u  \} ~,
\eq
while for $ZZ$ we have  both u-channel and t-channel top exchange leading to
\bqa
A^{ZZ}_u &\to & - {e^2(E_t+m_t)\over 2(u-m^2_t)}\{g^{2L}_{Zt}(T^{1L}_u+T^{2L}_u+T^{3L}_u)
\nonumber \\
&& +  g^{2R}_{Zt}(T^{1R}_u+T^{2R}_u+T^{3R}_u)+g^{L}_{Zt}g^{R}_{Zt}T^{0}_u\} ~~, \nonumber\\
A^{ZZ}_t &\to &
- {e^2(E_t+m_t)\over 2(t-m^2_t)}\{g^{2L}_{Zt}(T^{1L}_t+T^{2L}_t+T^{3L}_t) \nonumber \\
 && +  g^{2R}_{Zt}(T^{1R}_t+T^{2R}_t+T^{3R}_t)+g^{L}_{Zt}g^{R}_{Zt}T^{0}_t\} ~~.
\eqa
For both $W^-W^+$  and $ZZ$ cases these are expressed in terms of
\bqa
&& T^{1L}_u=-2(\epsilon^{\prime 0}p^0_t-\epsilon^{\prime 1}p_t\sin\theta-\epsilon^{\prime 3}p_t\cos\theta)
(1-2\tau r_t)(r_t+2\tau')\left (\epsilon^0\delta_{\tau\tau'}+\sum^3_{k=1}\epsilon^kS^k \right )
~, \nonumber \\
&& T^{1R}_u=-2(\epsilon^{\prime 0}p^0_t-\epsilon^{\prime 1}p_t\sin\theta-\epsilon^{\prime 3}p_t\cos\theta)
(1+2\tau r_t)(r_t-2\tau')\left (\epsilon^0\delta_{\tau\tau'}-\sum^3_{k=1}\epsilon^kS^k \right )
~, \nonumber \\
&&T^{2L}_u=m_t(Y_{11}-Y_{12})(1+2\tau r_t)(r_t+2\tau') ~~, \nonumber \\
&& T^{2R}_u=m_t(Y_{11}-Y_{12})(1-2\tau r_t)(r_t-2\tau') ~~, \nonumber \\
&& T^0_u=-m_t[Y_{11}r_t(1+4\tau\tau')-Y_{12}(2\tau'+2\tau r^2_t)]  ~~, \nonumber \\
&& T^{3L}_u=(1-2\tau r_t)(r_t+2\tau')(X_{11}-X_{12}) ~~, \nonumber \\
&& T^{3R}_u=(1+2\tau r_t)(r_t-2\tau')(X_{11}+X_{12}) ~~, \nonumber
\eqa
with
\bqa
 Y_{11} &= &\left (\epsilon^{\prime 0}\epsilon^{0}-\sum^3_{k=1}\epsilon^{\prime k}\epsilon^k \right )
 \delta_{\tau\tau'}
-i\Big [S^1(\epsilon^{\prime 2}\epsilon^{3}-\epsilon^{\prime 3}\epsilon^{2} )
+S^2(\epsilon^{\prime 3}\epsilon^{1}-\epsilon^{\prime 1}\epsilon^{3}) \nonumber \\
&+& S^3(\epsilon^{\prime 1}\epsilon^{2}-\epsilon^{\prime 2}\epsilon^{1}) \Big ] ~~, \nonumber \\
Y_{12} &= &S^1(\epsilon^{\prime 1}\epsilon^{0}-\epsilon^{\prime 0}\epsilon^{1})
+S^2(\epsilon^{\prime 2}\epsilon^{0}-\epsilon^{\prime 0}\epsilon^{2})
+S^3(\epsilon^{\prime 3}\epsilon^{0}-\epsilon^{\prime 0}\epsilon^{3}) ~~, \nonumber
\eqa
\bqa
&& X_{11}\mp X_{12} = l^0(\epsilon^{\prime 0}\epsilon^{0}+\sum^3_{k=1}\epsilon^{\prime k}\epsilon^k)
+l(\epsilon^{\prime 0}\epsilon^{3}+\epsilon^{\prime 3}\epsilon^0)
+il(\epsilon^{\prime 2}\epsilon^{1}-\epsilon^{\prime 1}\epsilon^2))\nonumber\\
&&+S^1[il(\epsilon^{\prime 2}\epsilon^{0}-\epsilon^{\prime 0}\epsilon^2)
+il^0(\epsilon^{\prime 2}\epsilon^{3}-\epsilon^{\prime 3}\epsilon^2)
\pm l^0(\epsilon^{\prime 0}\epsilon^{1}+\epsilon^{\prime 1}\epsilon^0)
\pm l(\epsilon^{\prime 1}\epsilon^{3}+\epsilon^{\prime 3}\epsilon^1)]
\nonumber\\
&&+S^2[il(\epsilon^{\prime 0}\epsilon^{1}-\epsilon^{\prime 1}\epsilon^0)
+il^0(\epsilon^{\prime 3}\epsilon^{1}-\epsilon^{\prime 1}\epsilon^3)
\pm l^0(\epsilon^{\prime 0}\epsilon^{2}+\epsilon^{\prime 2}\epsilon^0)
\pm l(\epsilon^{\prime 2}\epsilon^{3}+\epsilon^{\prime 3}\epsilon^2)]\nonumber\\
&&+S^3[il^0(\epsilon^{\prime 1}\epsilon^{2}-\epsilon^{\prime 2}\epsilon^1)
+l^0(\epsilon^{\prime 0}\epsilon^{3}+\epsilon^{\prime 3}\epsilon^0)
\pm l(\epsilon^{\prime 0}\epsilon^{0}-\sum^3_{k=1}\epsilon^{\prime k}\epsilon^k)
\pm l(\epsilon^{\prime3}\epsilon^{3}+\epsilon^3\epsilon^{\prime3})] ~~, \nonumber
\eqa
and
\bqa
&& S^1=\cos\theta(\delta_{\tau+}\delta_{\tau'-}+\delta_{\tau-}\delta_{\tau'+})
+\sin\theta(\delta_{\tau+}\delta_{\tau'+}-\delta_{\tau-}\delta_{\tau'-}) ~~, \nonumber \\
&& S^2=-i(\delta_{\tau+}\delta_{\tau'-}-\delta_{\tau-}\delta_{\tau'+}) ~~, \nonumber \\
&& S^3=-\sin\theta(\delta_{\tau+}\delta_{\tau'-}+\delta_{\tau-}\delta_{\tau'+})
+\cos\theta(\delta_{\tau+}\delta_{\tau'+}-\delta_{\tau-}\delta_{\tau'-}) ~~. \nonumber
\eqa

For the t-channel contribution the $T^i_t$ terms are obtained from the $T^i_u$ ones,
 by the interchanges
 \[
u\to t ~~,~~  \epsilon\to \epsilon' ~~,~~  l\to -l ~~.
\]

\item

{\bf  s-channel $V=\gamma,Z$ exchange, only for $WW$. }
\bqa
&&A^{WW}_{s,V}\to ~{e^2g_V(E_t+m_t)\over 2(s-m^2_V+im_V\Gamma_V)}
\Bigg \{-~{4ll^0\over m_W}\delta_{\lambda'0} \left [\epsilon^0
\delta_{\tau\tau'}+\sum^3_{k=1}\epsilon^kS^k \right ]
+~{4ll^0\over m_W}\delta_{\lambda 0}\Big [\epsilon^{\prime0}
\delta_{\tau\tau'}\nonumber\\
&&+\sum^3_{k=1}\epsilon^{\prime k}S^k \Big ]-2l\epsilon.\epsilon'S^3  \Bigg  \}.
\left \{g^L_{Vt}(1-2\tau r_t)(r_t+2\tau')+g^R_{Vt}(1+2\tau r_t)(r_t-2\tau')\right \} ~. \nonumber
\eqa

\item

{\bf  s-channel $H$ exchange, for both $W^-W^+$ or $ZZ$}
\[
A^{XX}_{s,H}\to ~{2e^2g_{HXX}g_{Htt}p_t\over s-m^2_H+im_H\Gamma_H}(\epsilon.\epsilon')
\delta_{\tau\tau'} ~,
\]
with $g_{HXX}=g_{HWW},g_{HZZ}$.

\end{itemize}

\begin{figure}[p]
\[
\epsfig{file=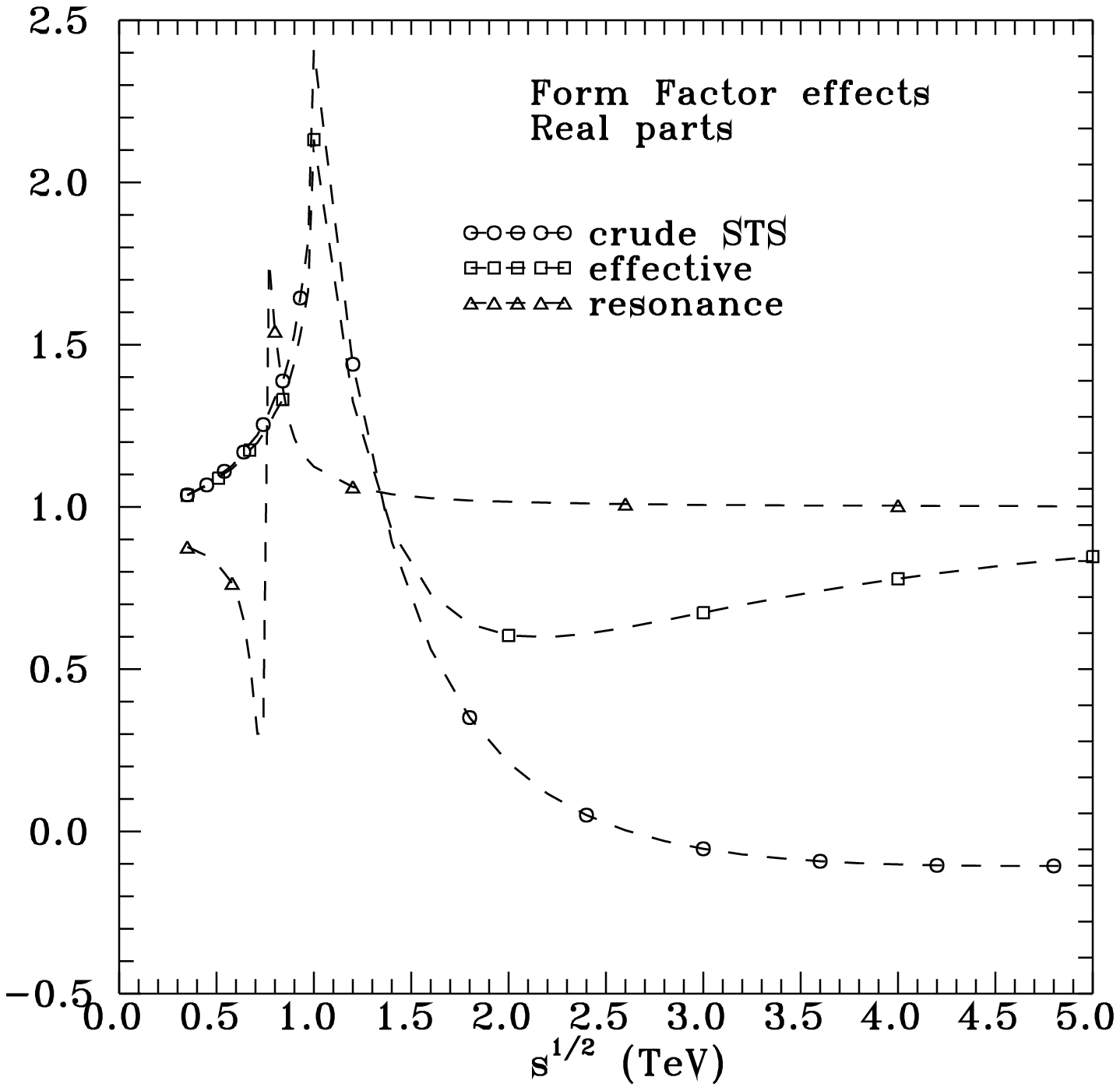, height=6.5cm}\hspace{0.5cm}
\epsfig{file=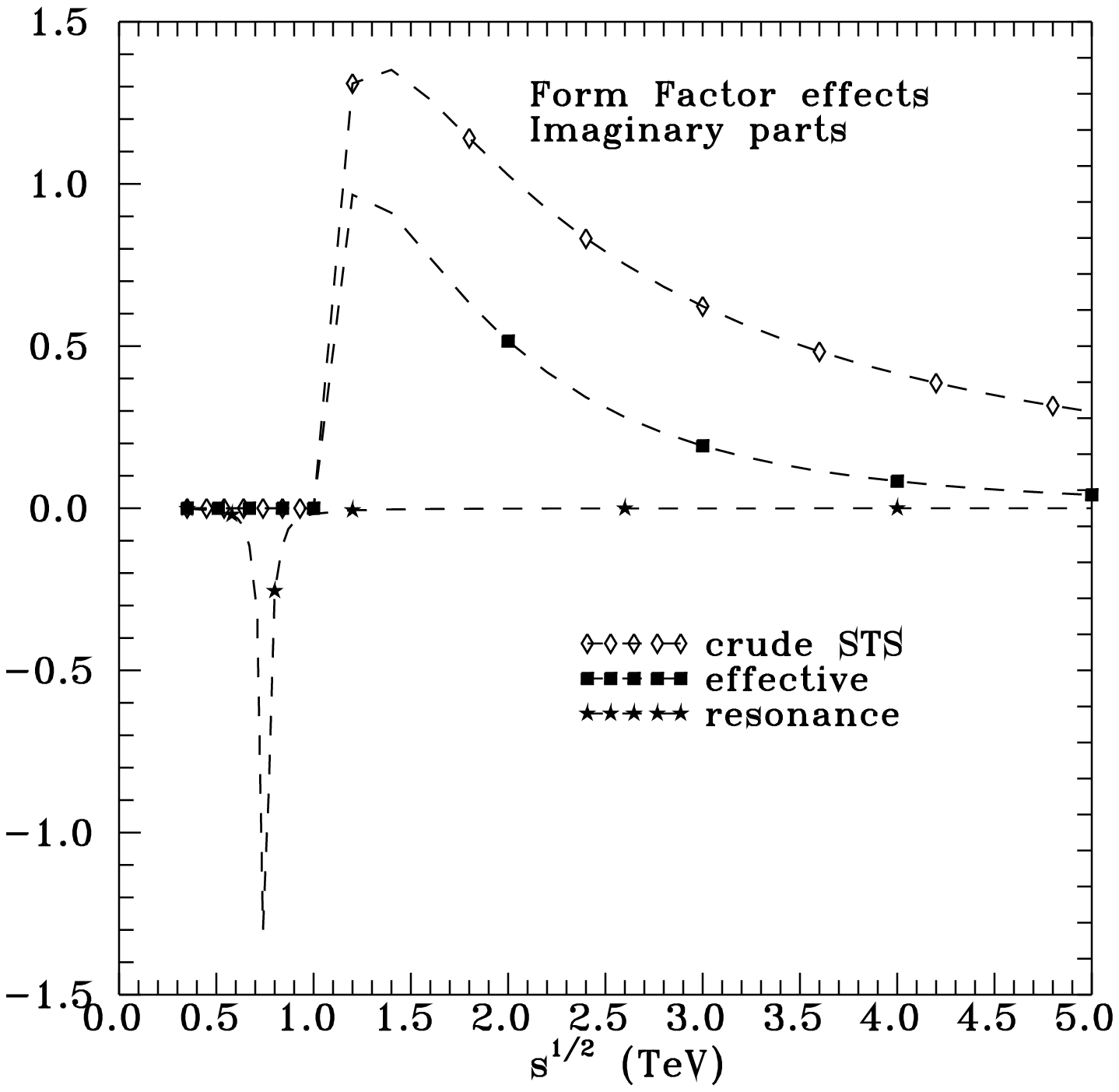,height=6.5cm}
\]
\vspace{-0.8cm}
\caption[1]{The shapes of the real parts (left panel) and imaginary parts (right panel) of the
$Ht\bar t$  form factors; the {\it crude} lines are
based on (\ref{crude}) and (\ref{STS}), the {\it effective} lines on (\ref{effective}) and (\ref{STS}) and the
 {\it resonance} lines are based on (\ref{resonance}).}
\label{Fg1}
\end{figure}

\clearpage

\begin{figure}[p]
\[
\epsfig{file=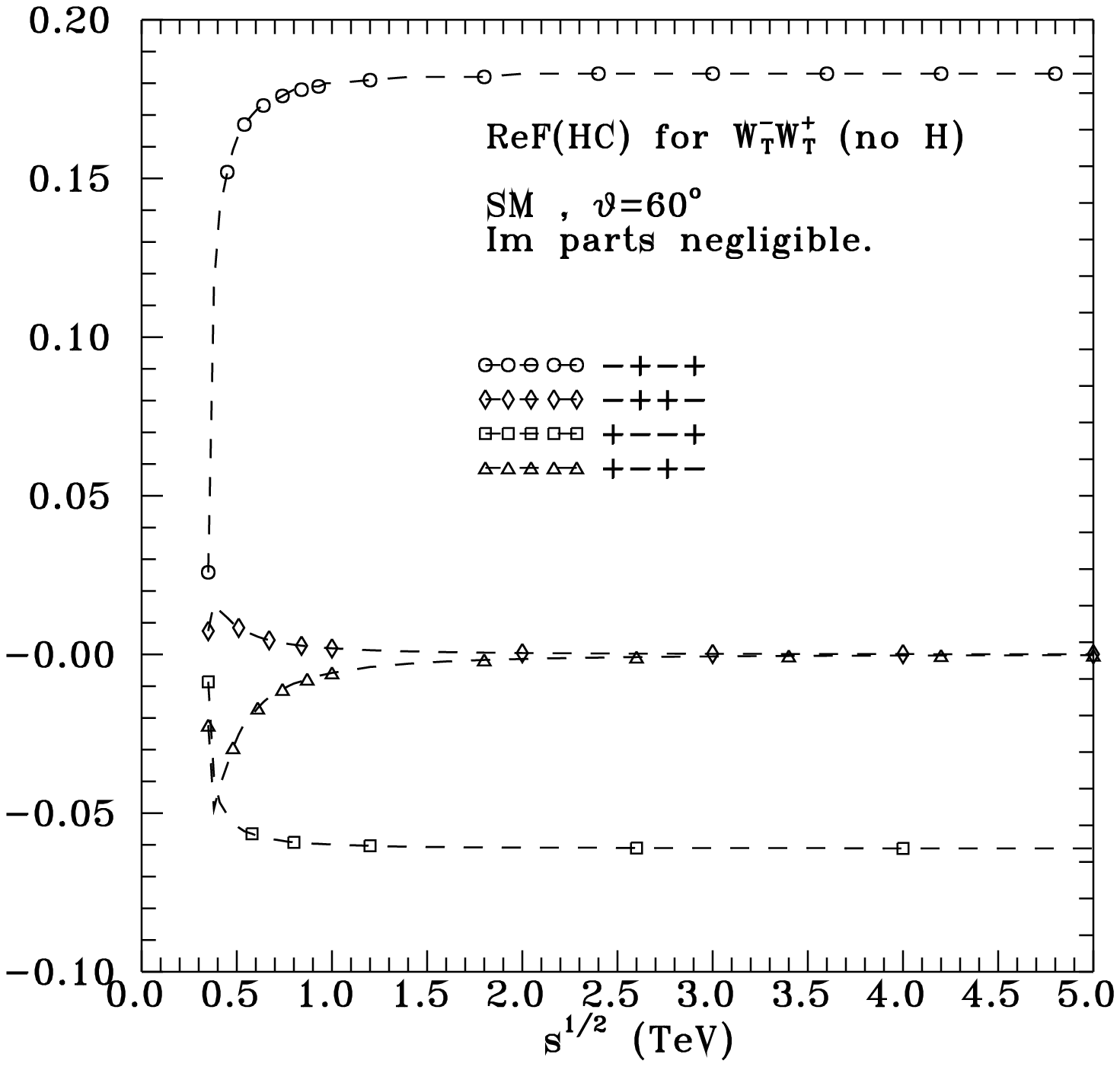, height=6.5cm}\hspace{0.5cm}
\epsfig{file=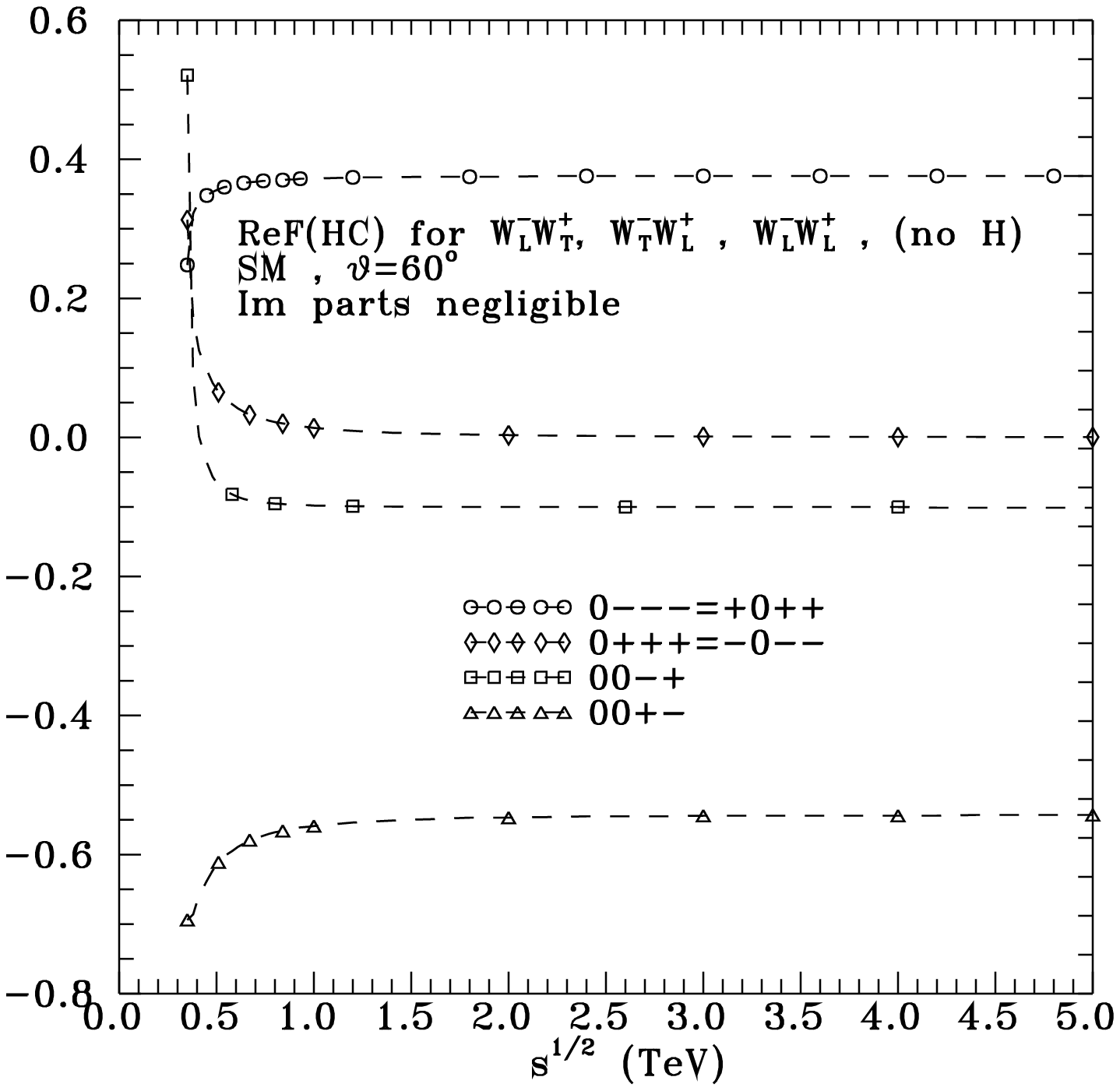,height=6.5cm}
\]
\[
\epsfig{file=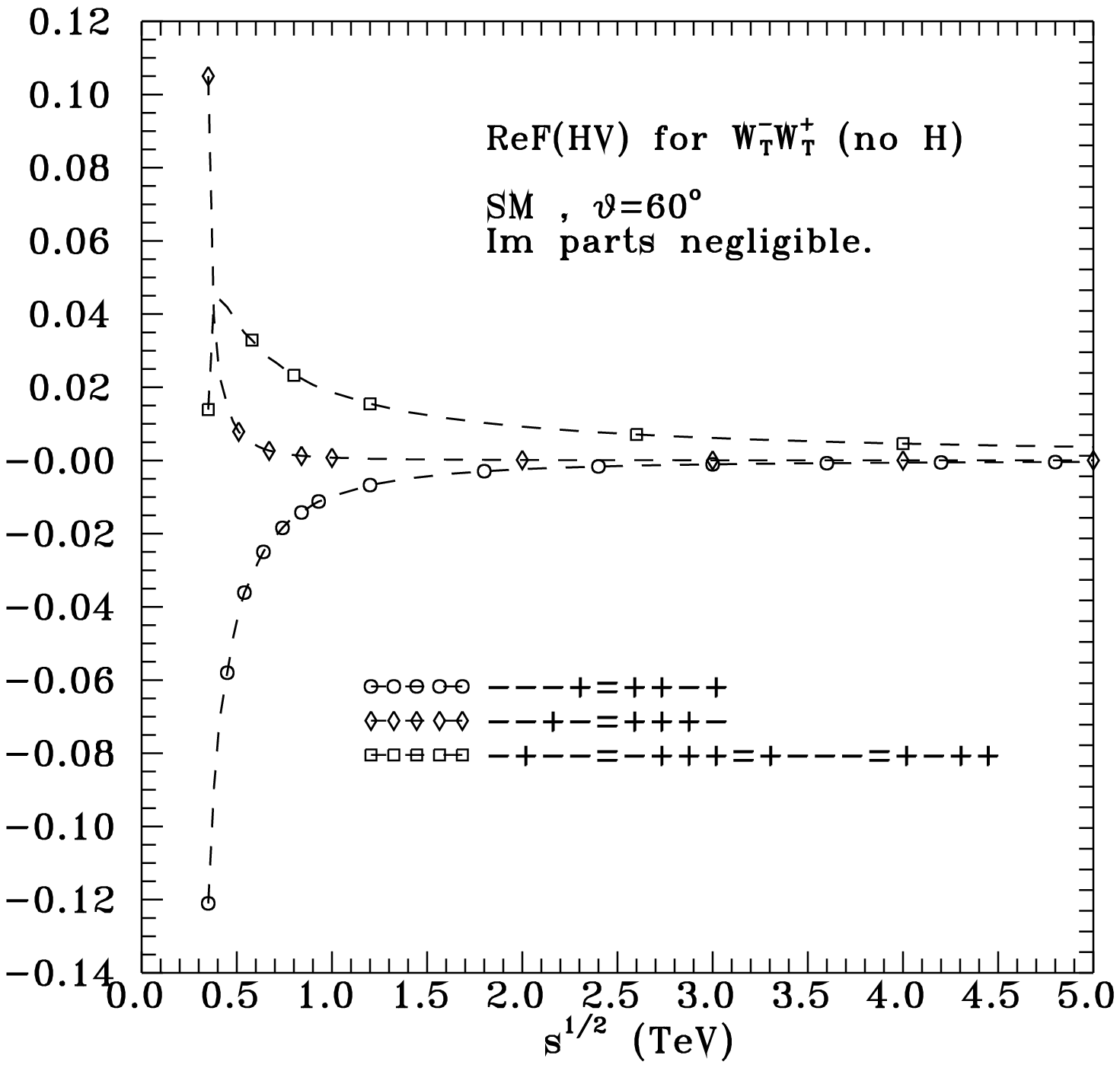, height=6.5cm}\hspace{0.5cm}
\epsfig{file=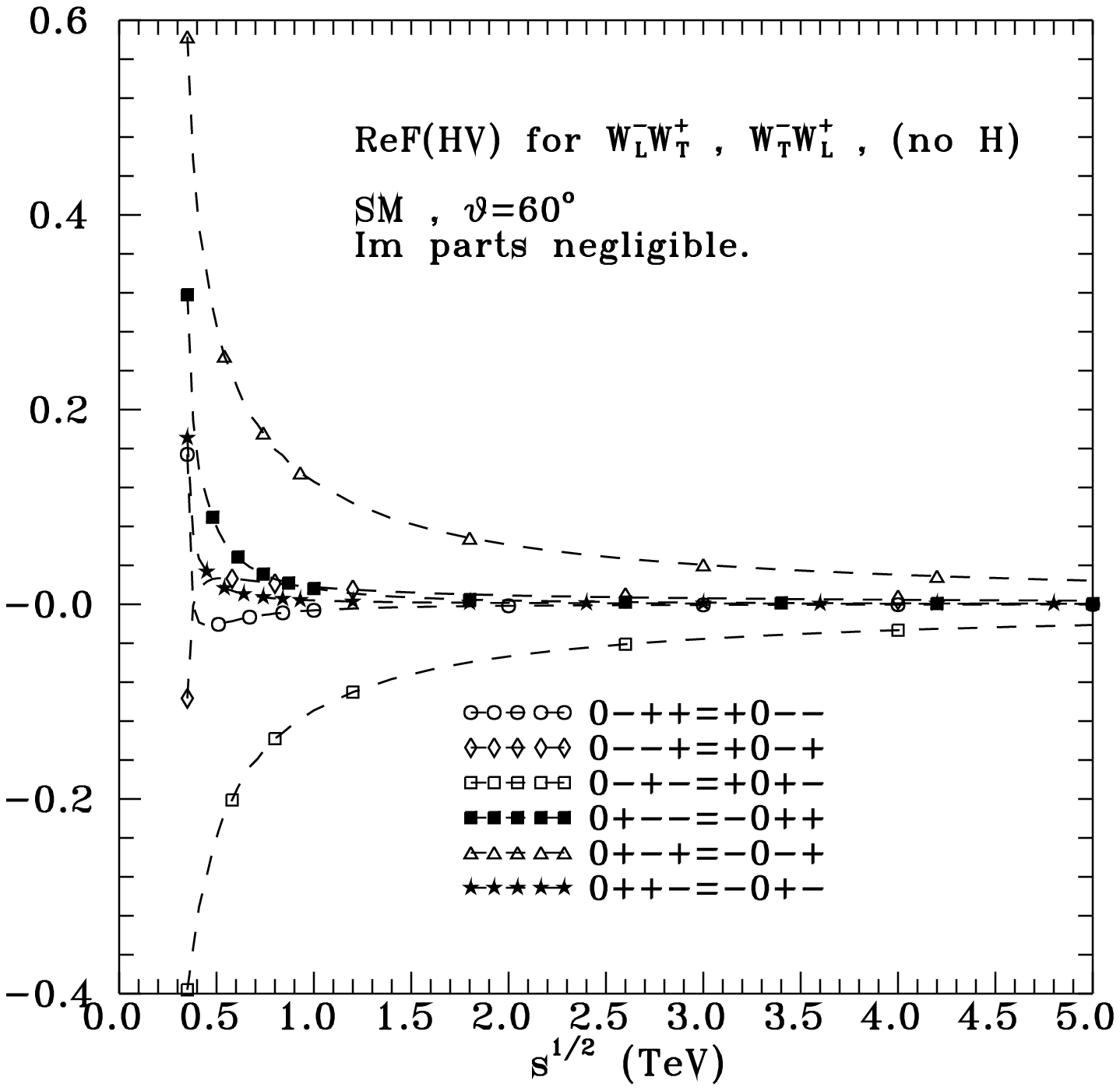, height=6.5cm}
\]
\caption[1]{Energy dependence of the H-insensitive tree-level SM amplitudes for
$W^-_\lambda W^+_{\lambda'} \to t_\tau \bar t_{\tau'}$. Upper panels are for the  HC amplitudes of
(\ref{WW-HC-SMamp}), and lower panels are for the HV ones of (\ref{WW-HV1-SMamp}) at $\theta=60^\circ$.
Left panels present TT amplitudes, while  right panels involve amplitudes
containing at least one longitudinal $W$. }
\label{Fg2}
\end{figure}

\clearpage

\begin{figure}[p]
\vspace{-0.5cm}
\[
\epsfig{file=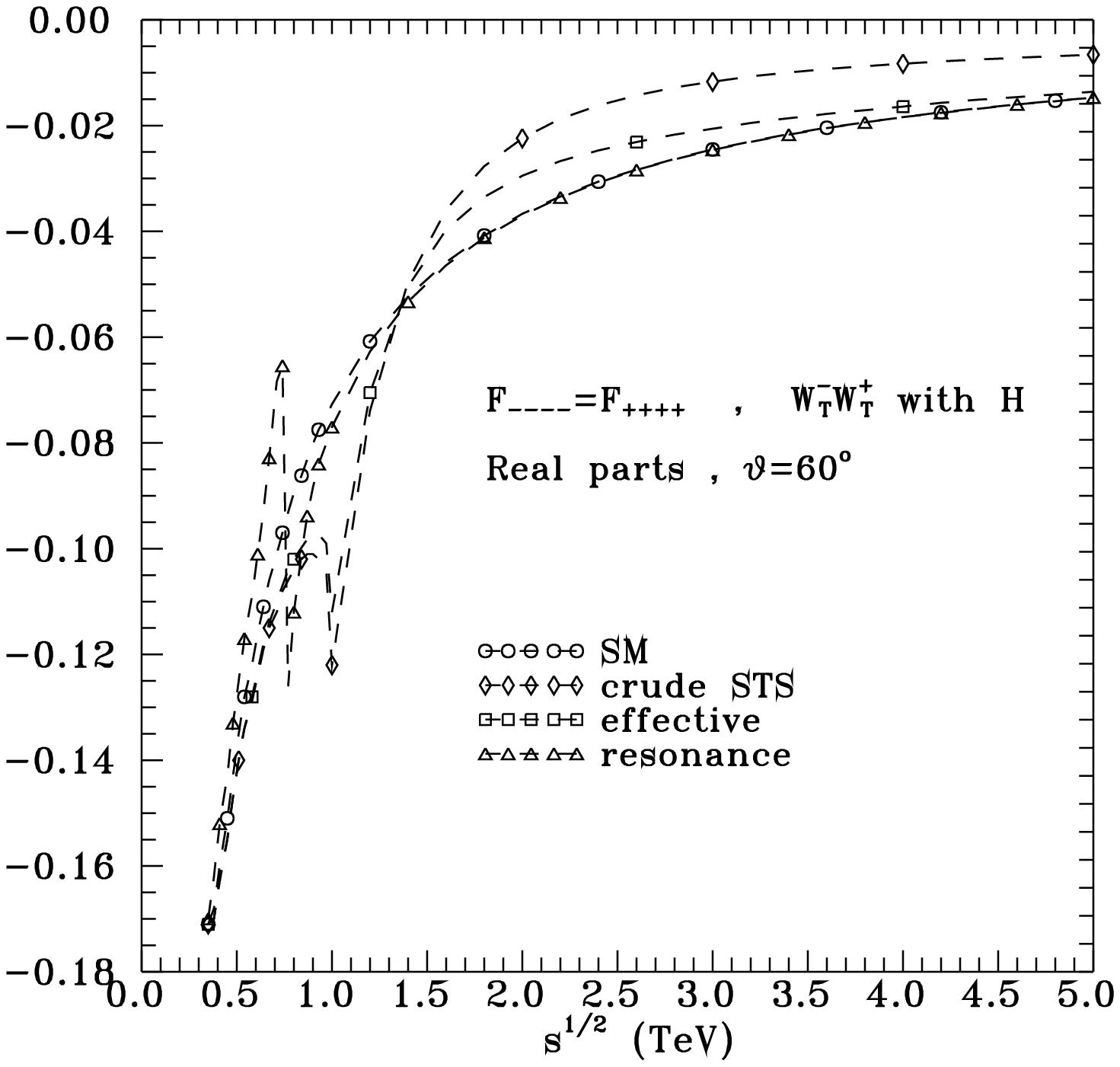, height=6.cm}\hspace{0.5cm}
\epsfig{file=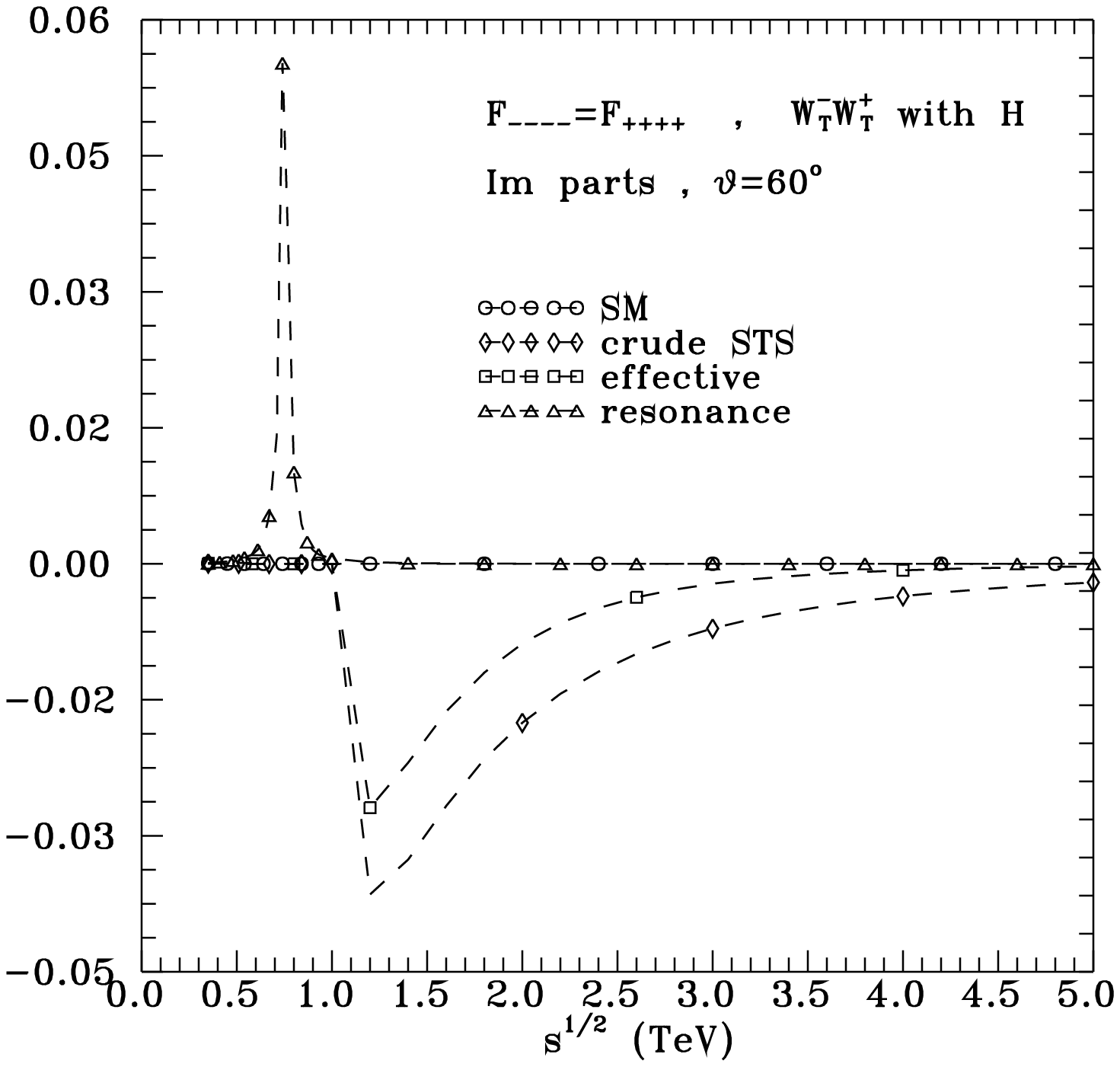,height=6.cm}
\]
\[
\epsfig{file=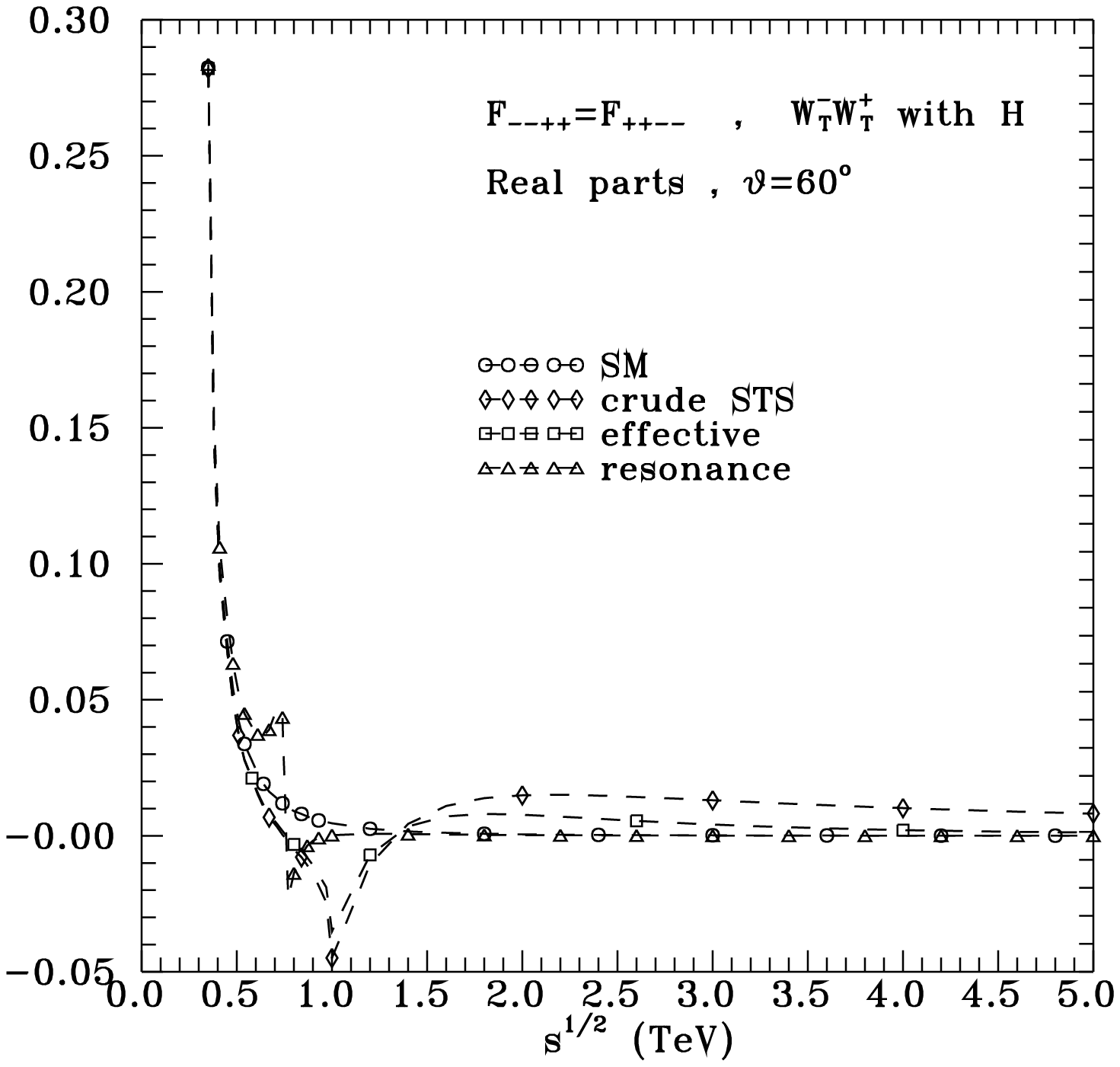, height=6.cm}\hspace{0.5cm}
\epsfig{file=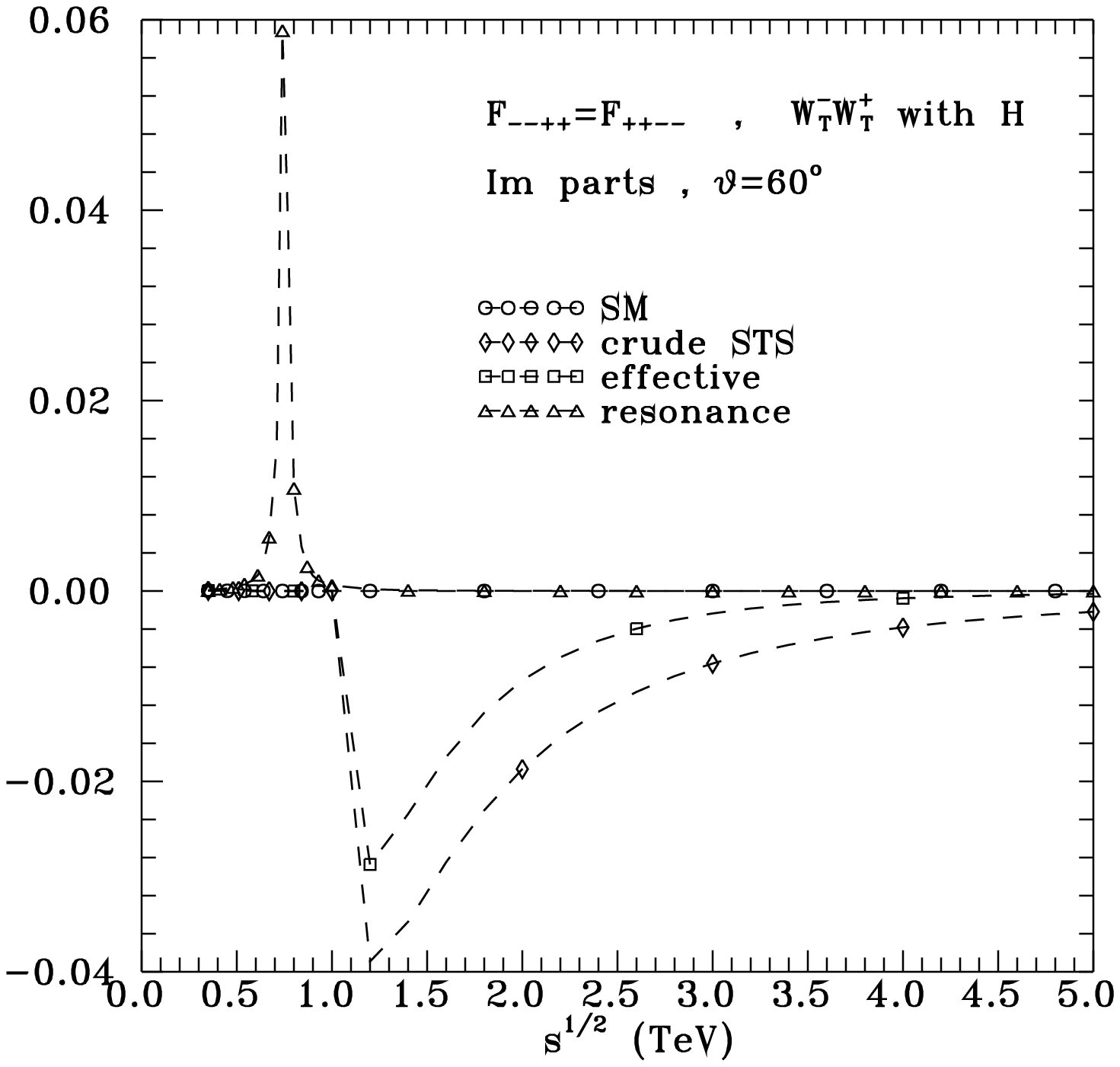,height=6.cm}
\]
\[
\epsfig{file=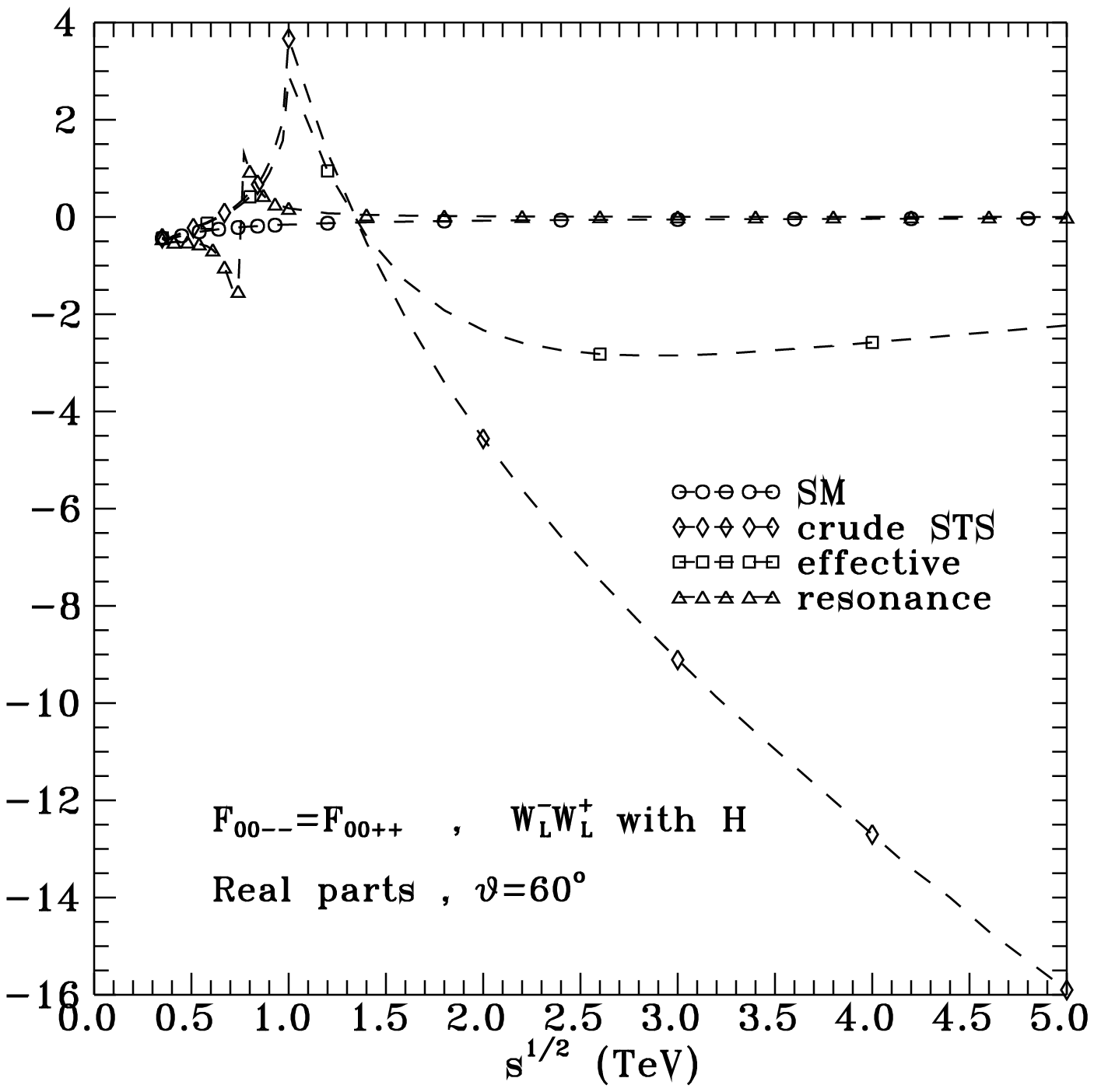, height=6.cm}\hspace{0.5cm}
\epsfig{file=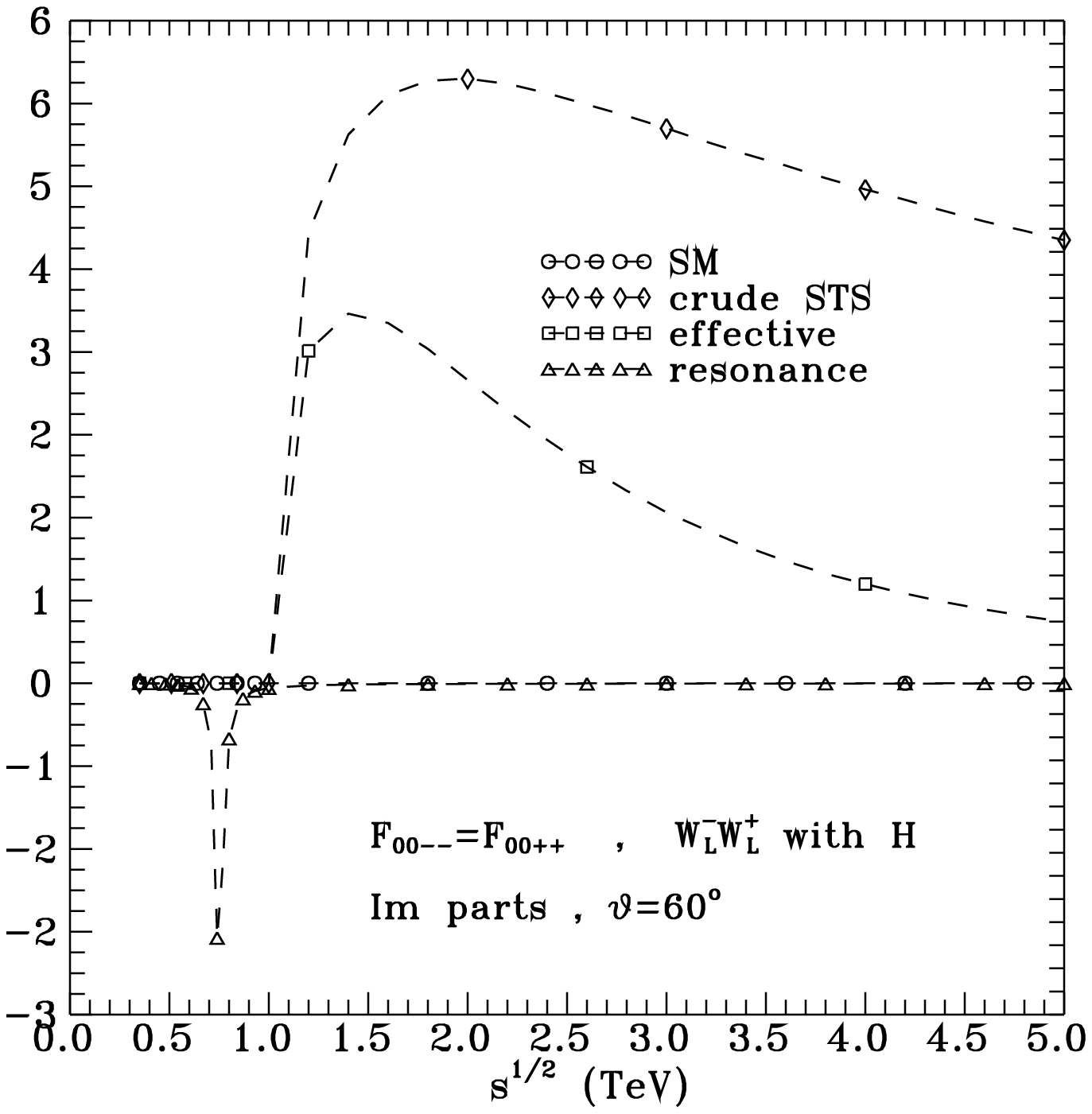,height=6.cm}
\]
\caption[1]{The SM results and the effects of the anomalous $Htt$ form factor
on the form factor sensitive  amplitudes
$W^-_\lambda W^+_{\lambda'} \to t_\tau \bar t_{\tau'}$ listed in (\ref{WW-HV2-SMamp}).
real (imaginary) parts are shown in the left (right) panels respectively.
The definition of {\it crude STS, effective} and
{\it resonance} form factor models, as in Fig.1.}
\label{Fg3}
\end{figure}

\clearpage

\begin{figure}[p]
\vspace{-0.5cm}
\[
\epsfig{file=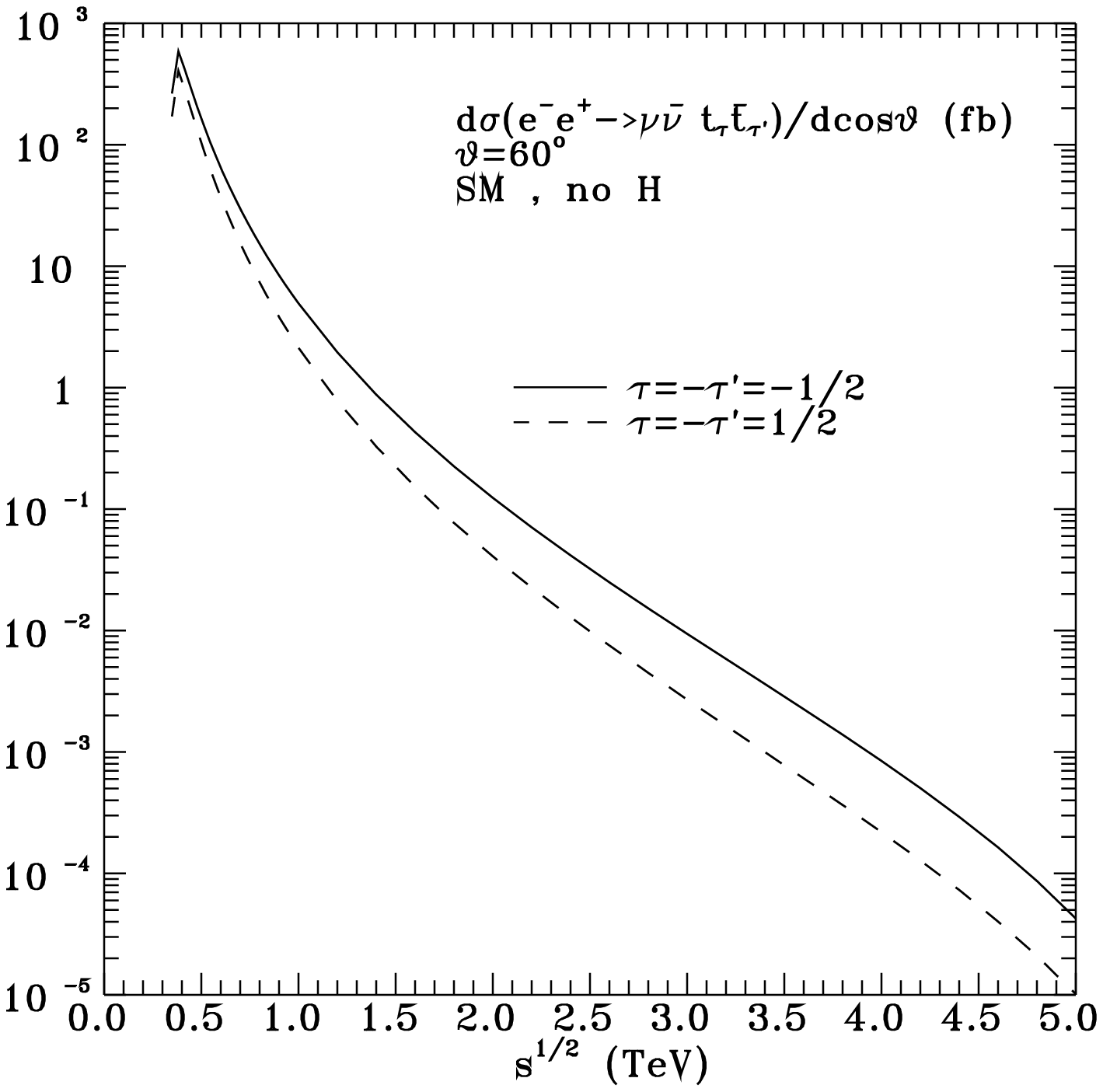,height=6.cm}
\]
\[
\epsfig{file=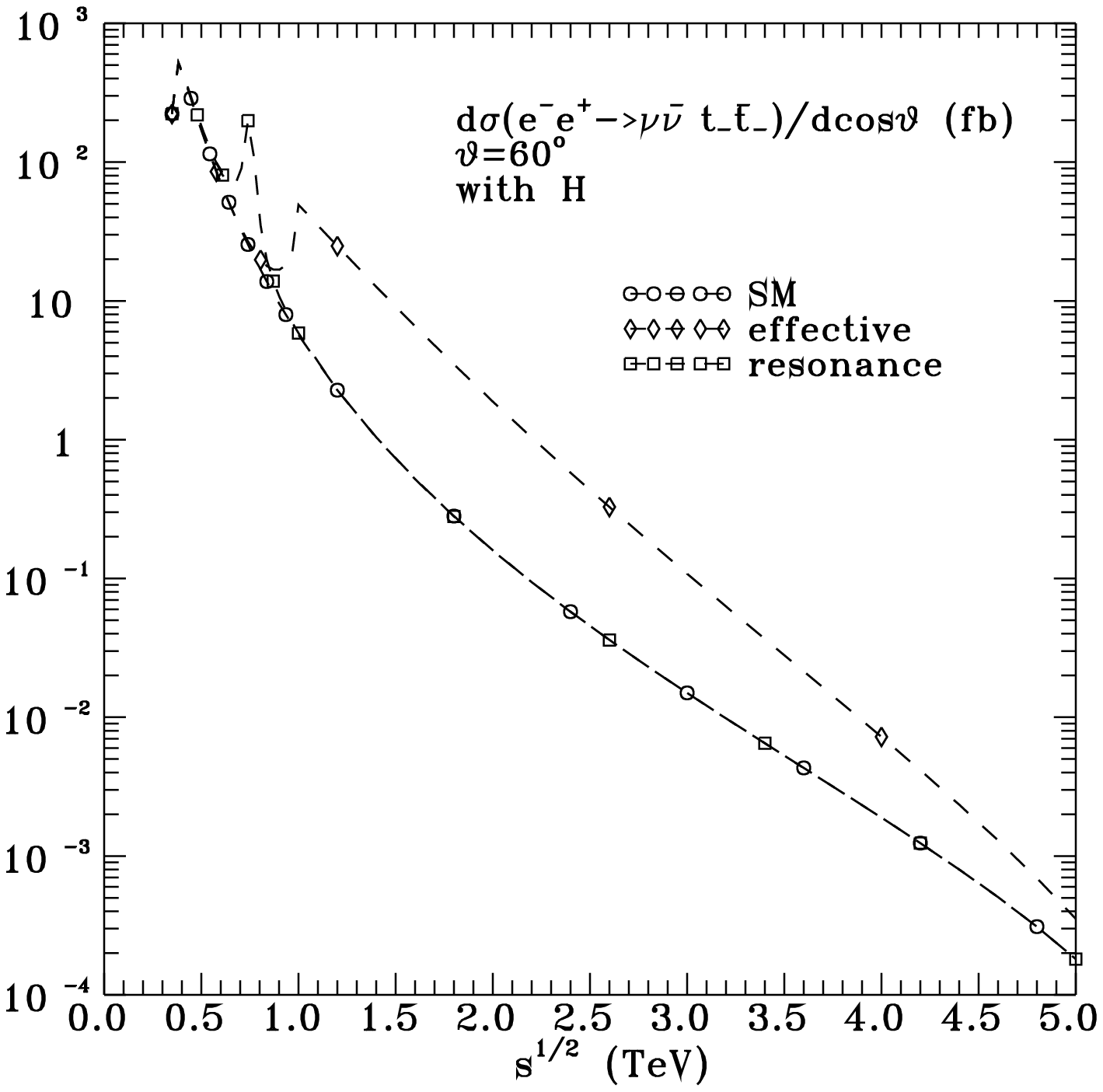, height=6.cm}\hspace{0.5cm}
\epsfig{file=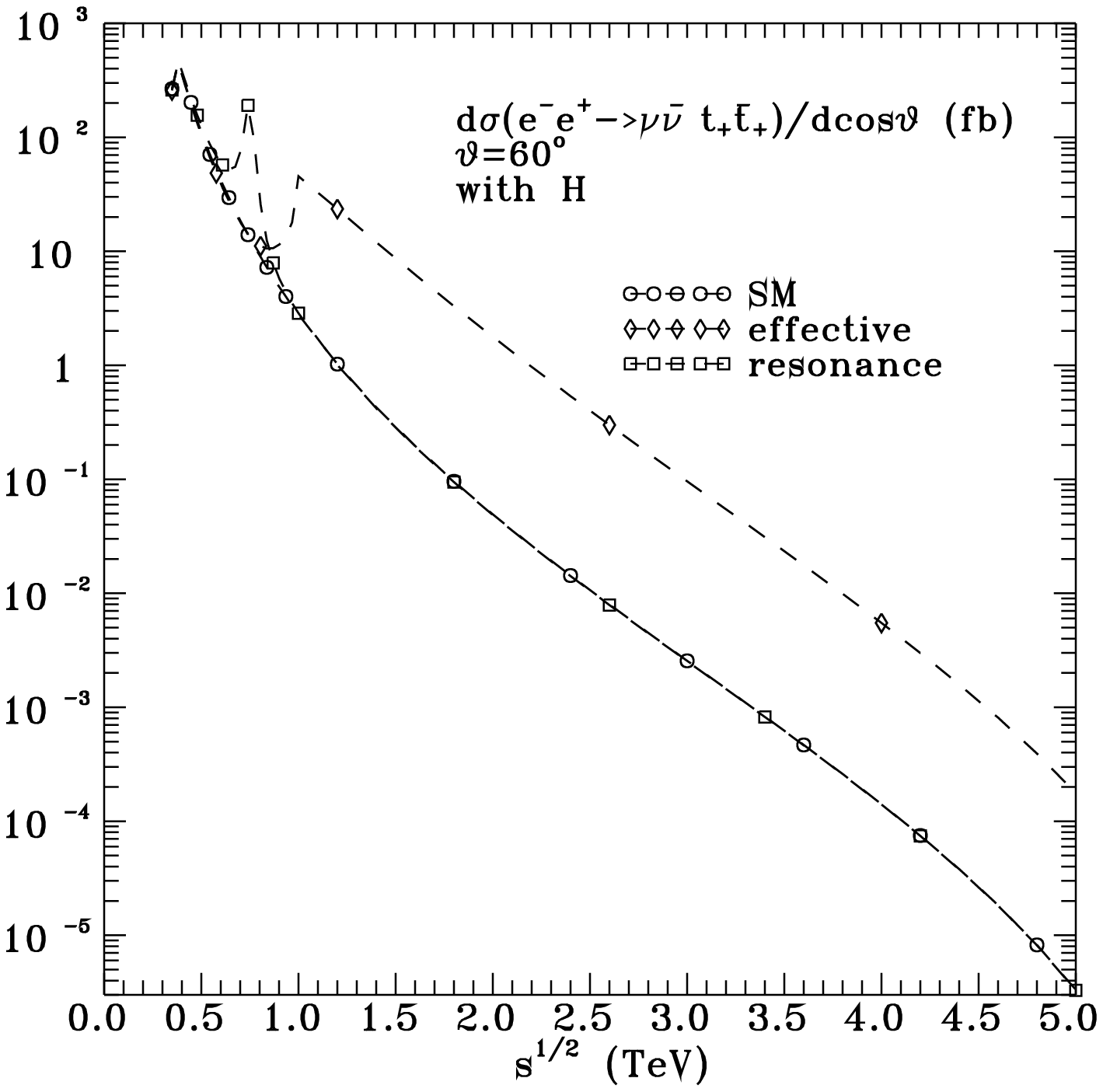,height=6.cm}
\]
\[
\epsfig{file=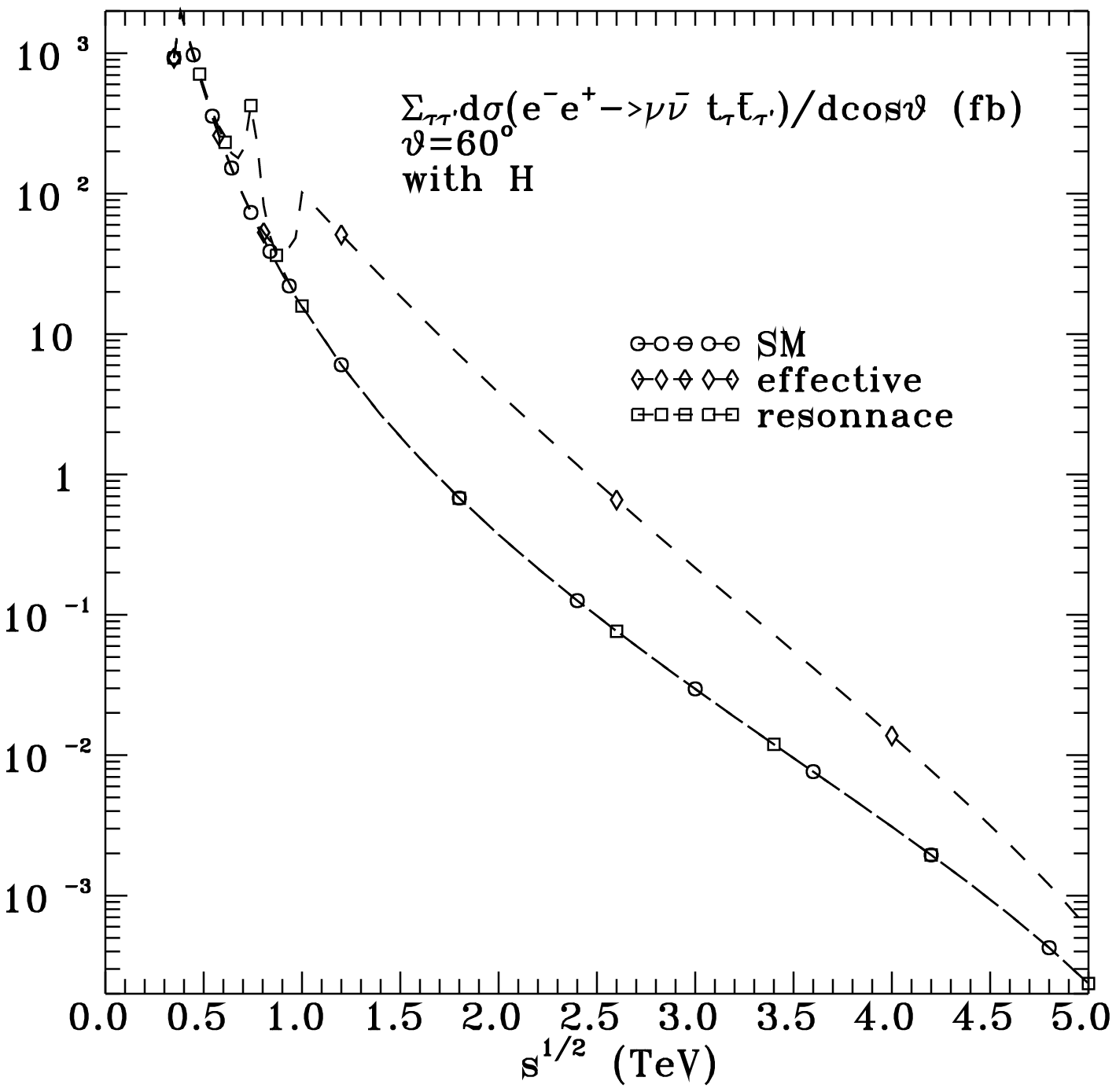, height=6.cm}\hspace{0.5cm}
\epsfig{file=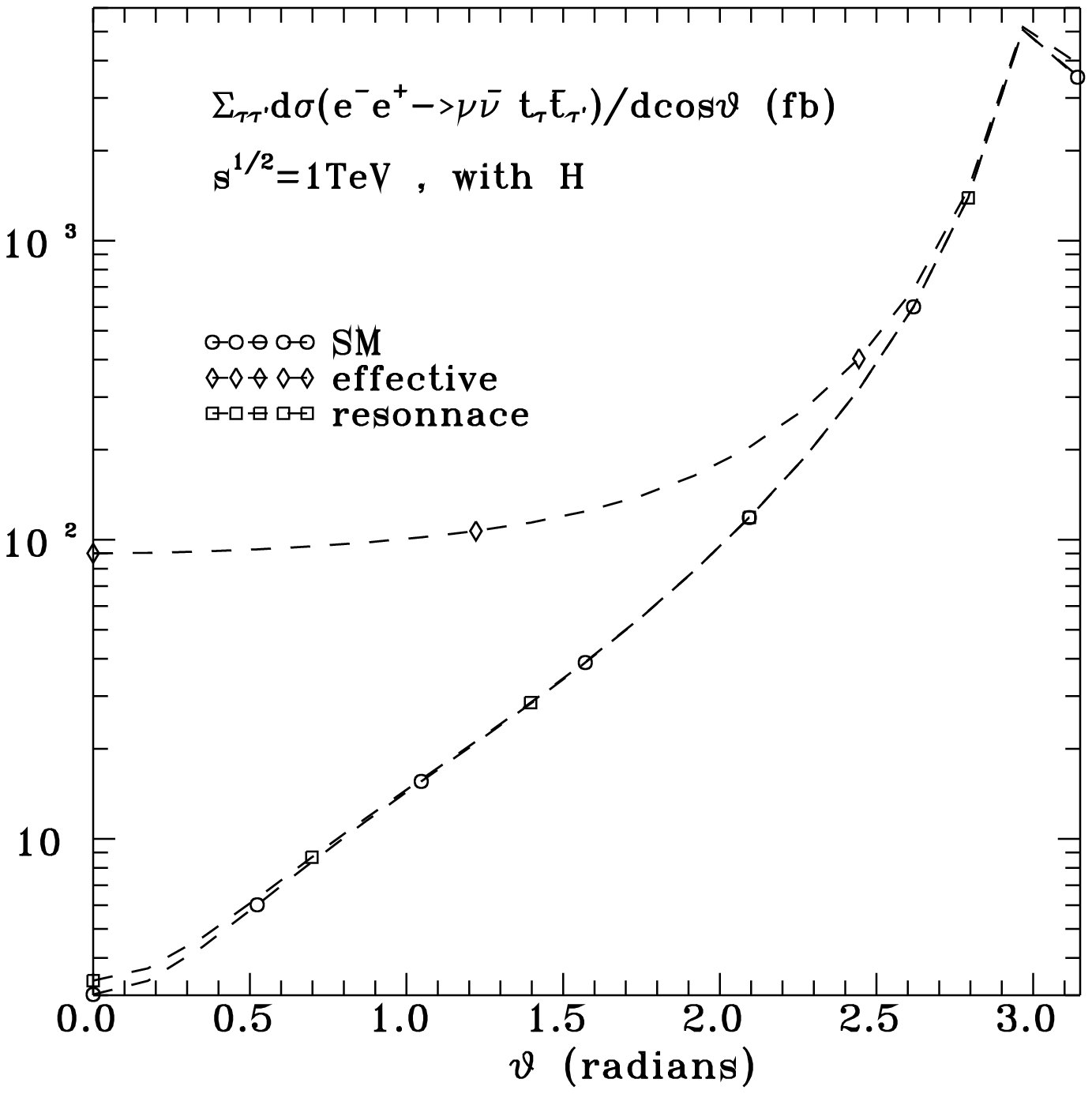, height=6.cm}
\]
\caption[1]{$d\sigma (e^-e^+ \to \nu \nu' t_\tau \bar t_{\tau '})/d\cos\theta$
cross sections.
The upper  panel gives energy dependencies for  (left-right) and (right-left) $t \bar t$ helicities,
while middle panels for (left-left) and (right-right) $t \bar t$ helicities, always at
$\theta=60^\circ$. The lower panels give energy (at $\theta=60^\circ$) and angular (at $\sqrt{s}=1$TeV)
dependencies when $t\bar t$-helicities  are not observed. The contribution from the $H$ form factors
of the three cases of Fig.1, exist only for the $t\bar t$ polarizations contributing
to the middle and lower panels.}
\label{Fg4}
\end{figure}

\clearpage

\begin{figure}[p]
\[
\epsfig{file=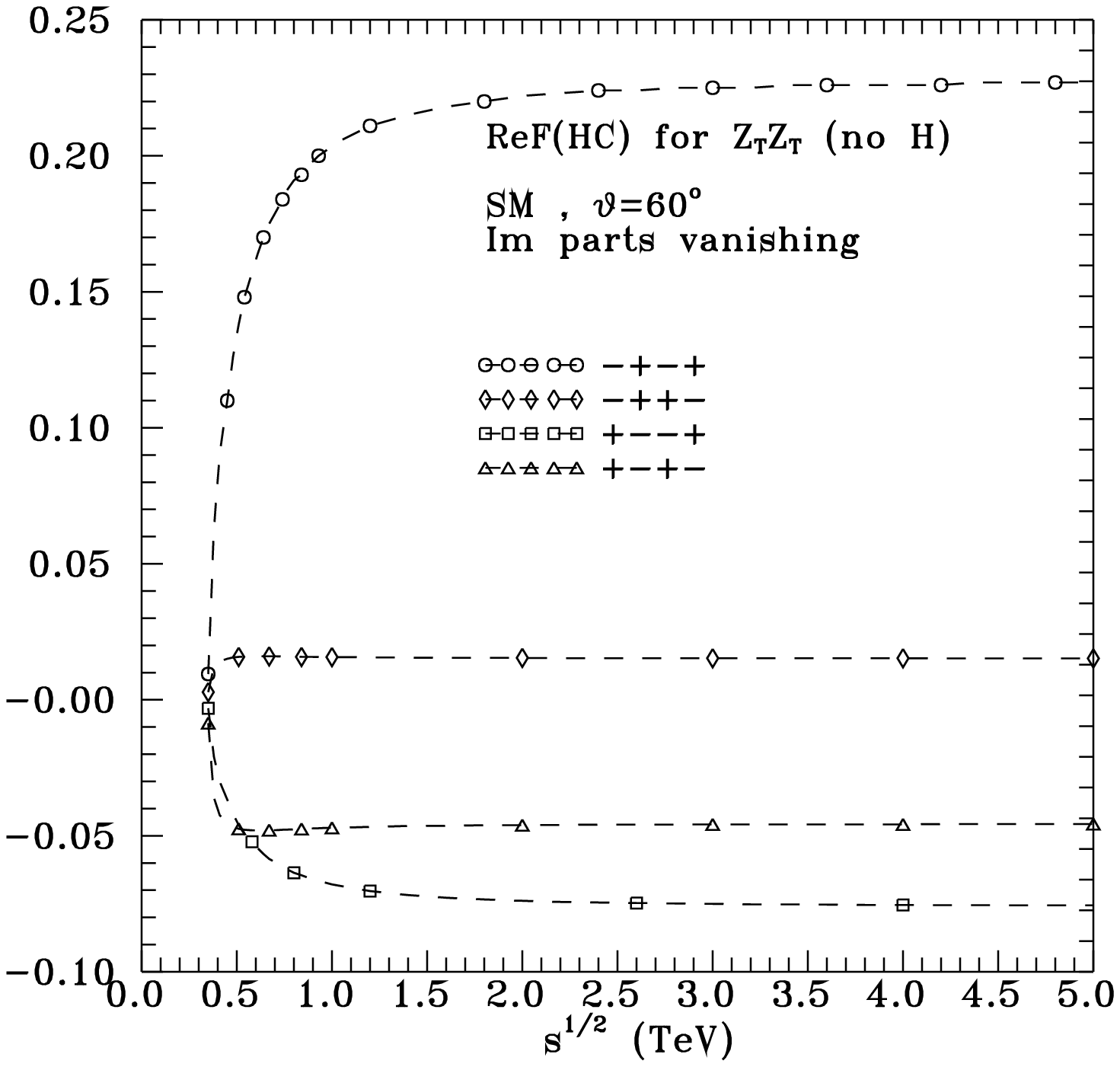, height=6.5cm}\hspace{0.5cm}
\epsfig{file=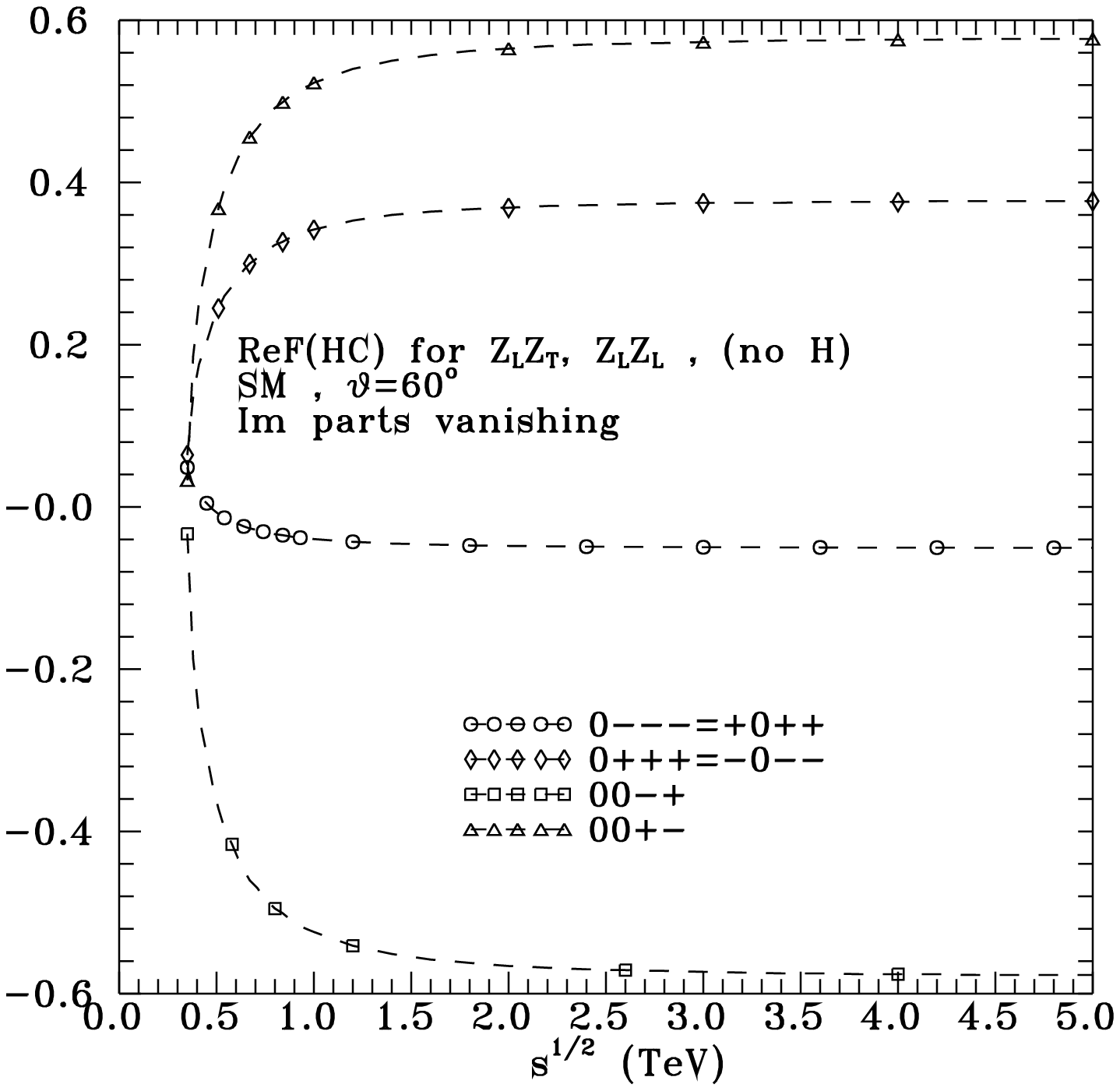, height=6.5cm}
\]
\[
\epsfig{file=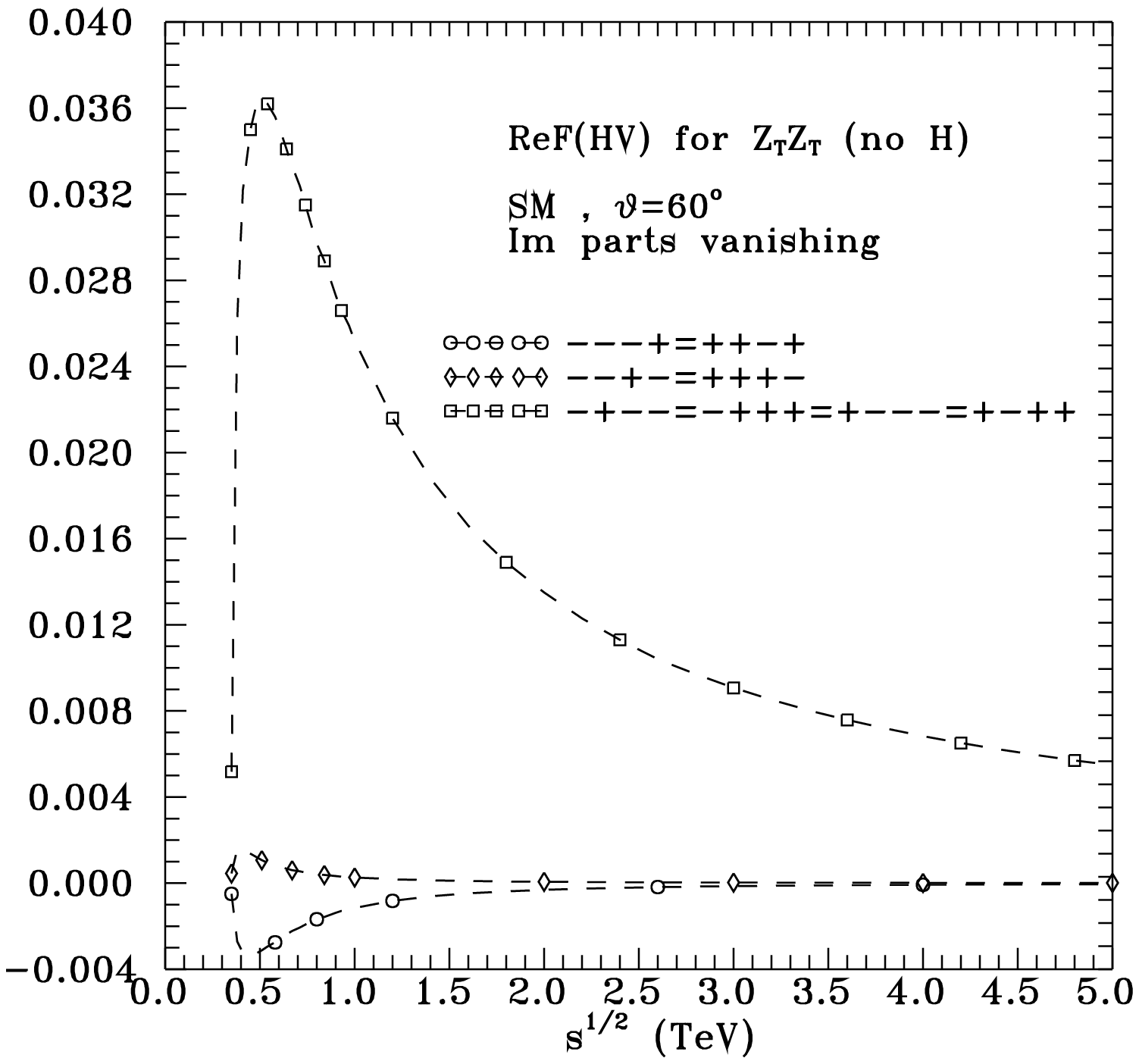, height=6.5cm}\hspace{0.5cm}
\epsfig{file=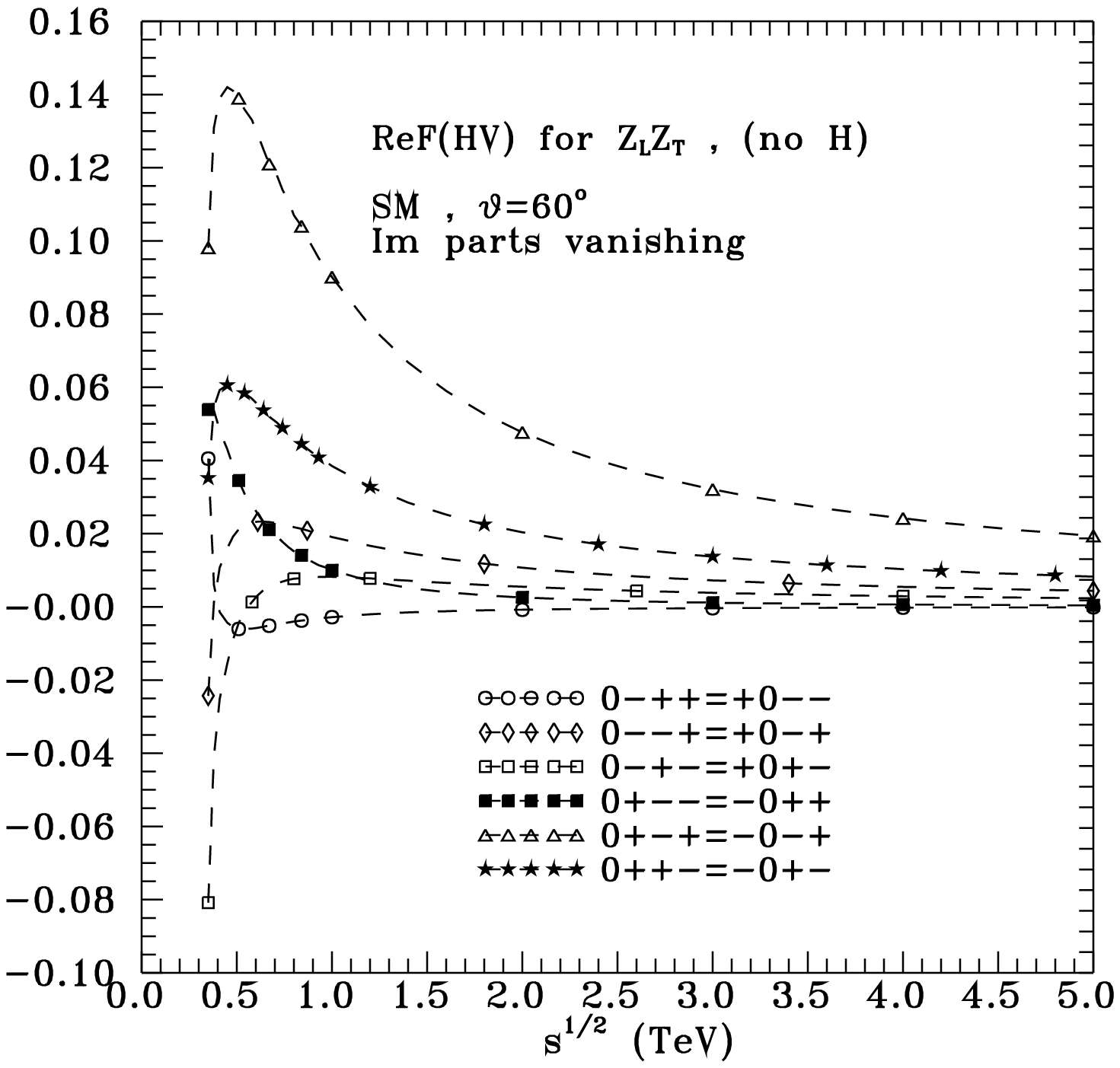,height=6.5cm}
\]
\caption[1]{ Energy dependence of the H-insensitive tree-level SM amplitudes for
$Z_\lambda Z_{\lambda'} \to t_\tau \bar t_{\tau'}$  at $\theta=60^\circ$.
Upper panels are for the HC amplitudes in (\ref{ZZ-HC-SMamp}) and lower panels are for
the HV ones in (\ref{ZZ-HV1-SMamp}).}
\label{Fg5}
\end{figure}

\clearpage

\begin{figure}[p]
\[
\epsfig{file=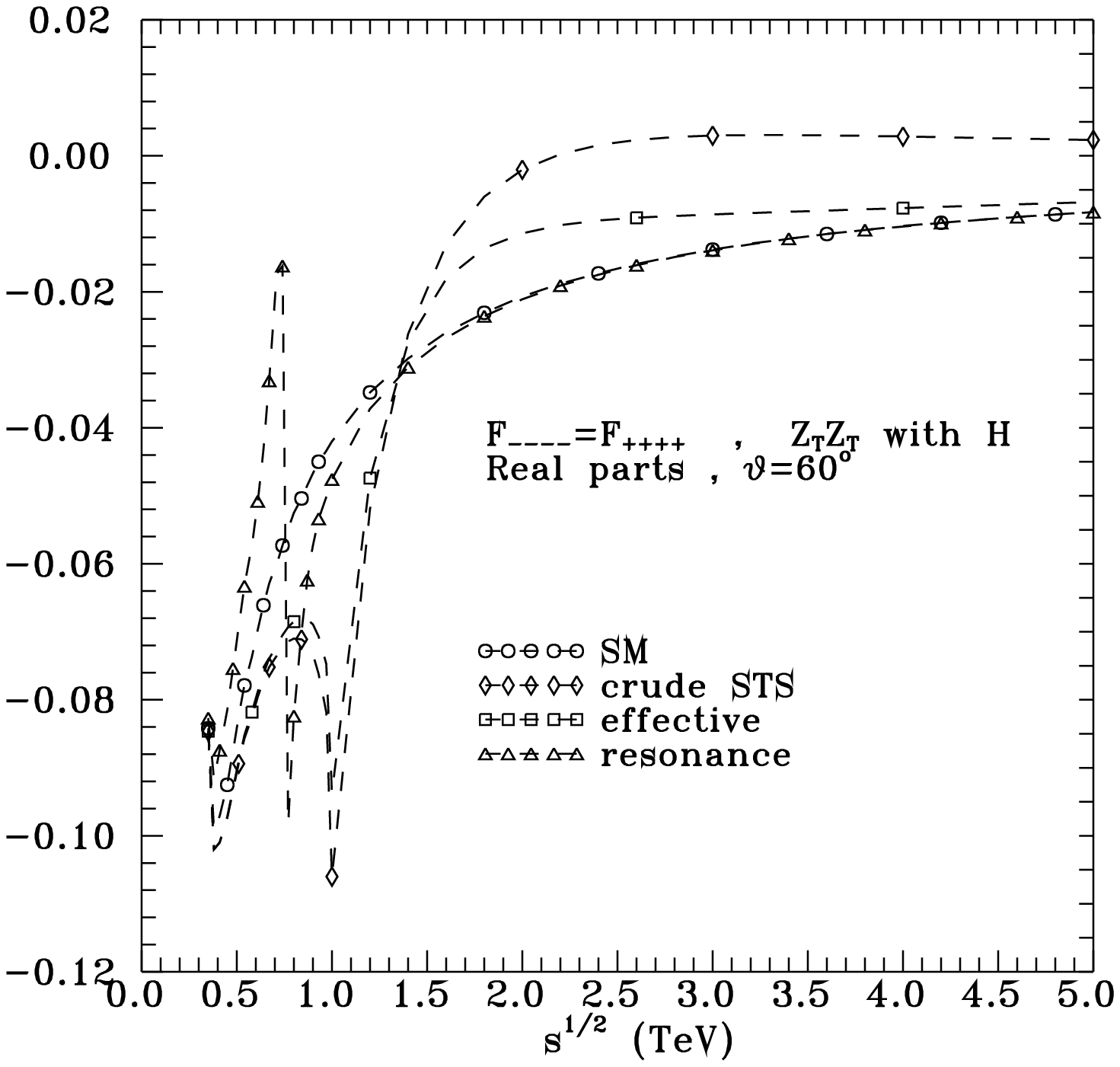, height=6.cm}\hspace{0.5cm}
\epsfig{file=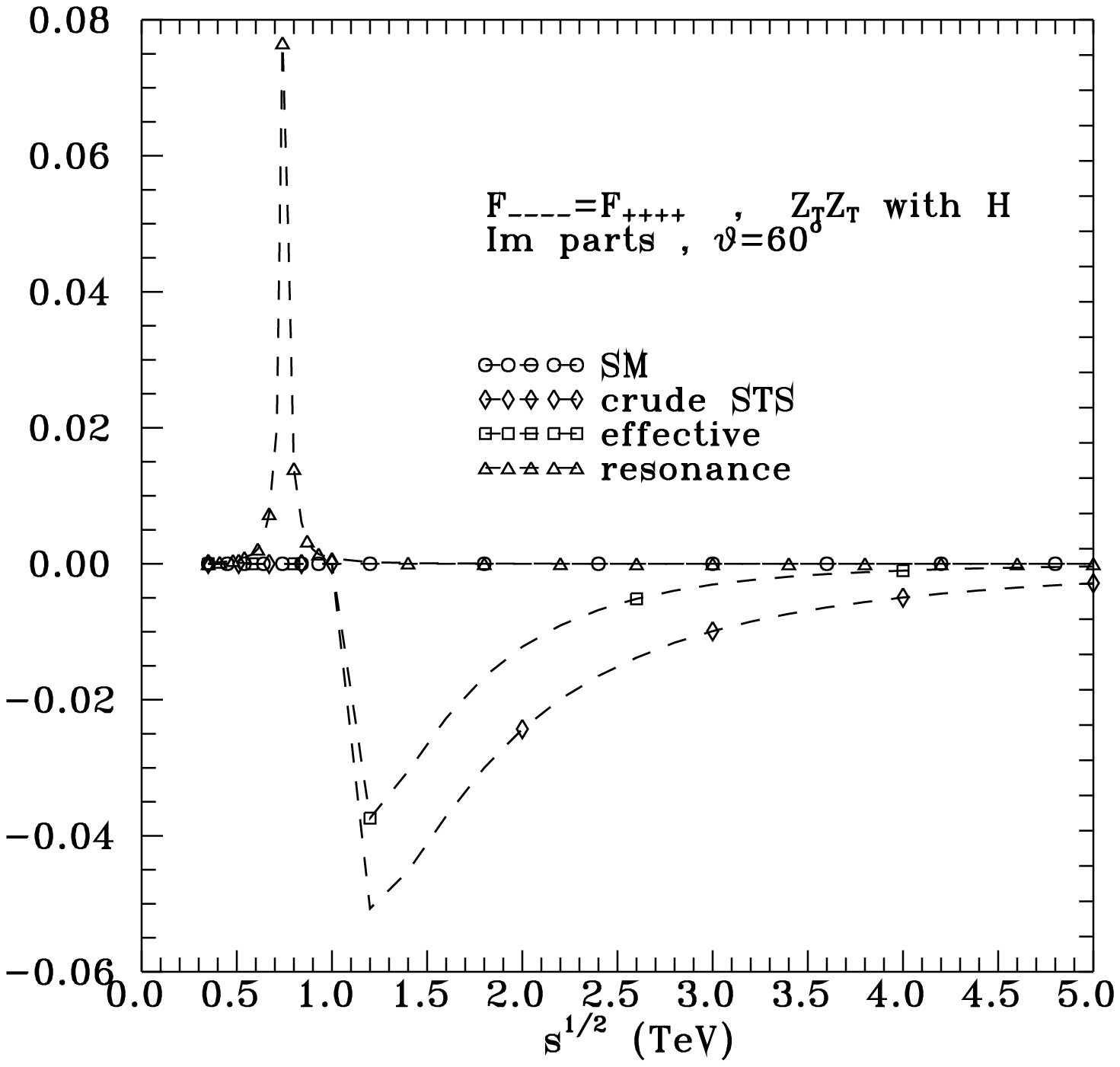, height=6.cm}
\]
\[
\epsfig{file=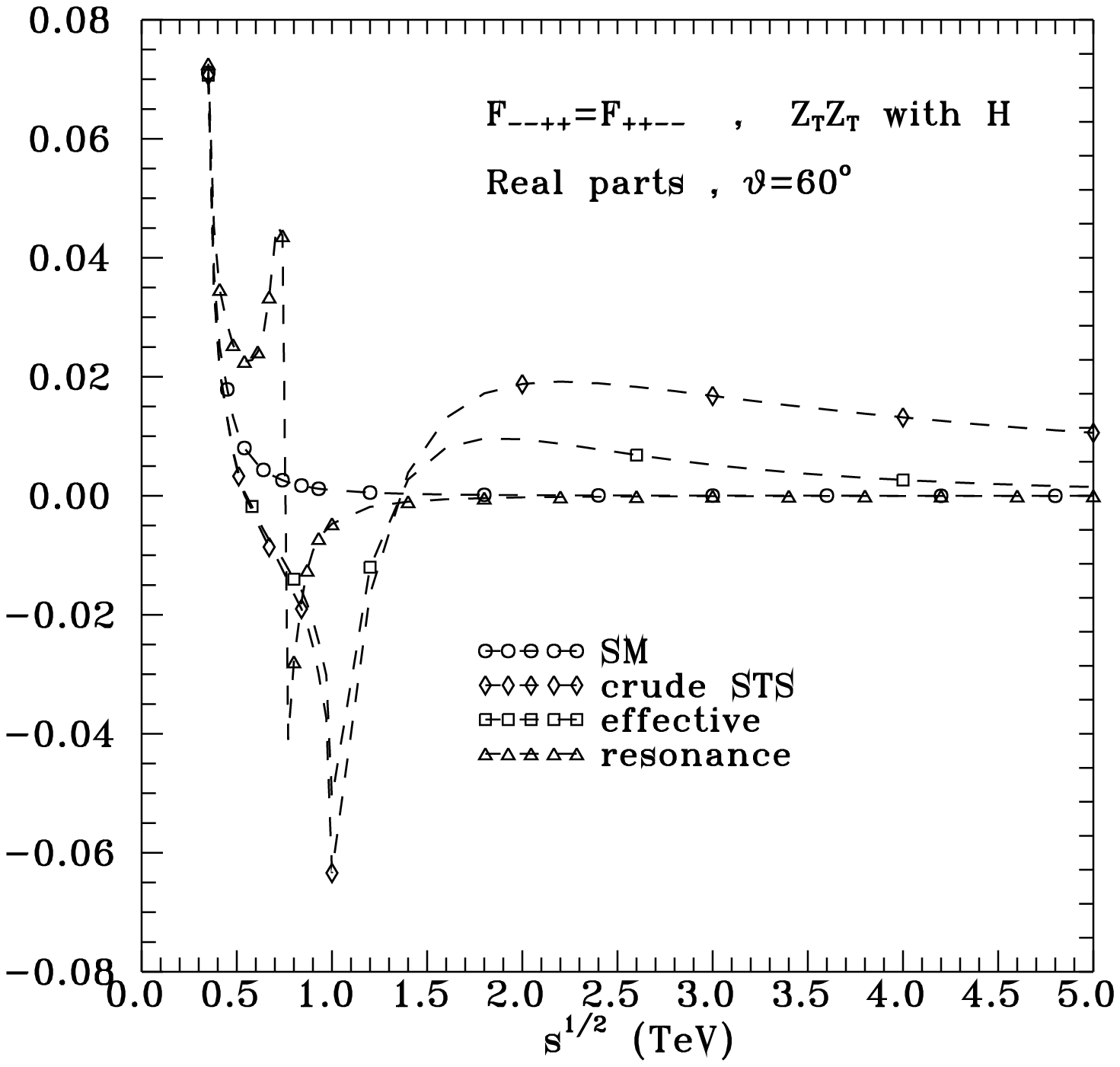, height=6.cm}\hspace{0.5cm}
\epsfig{file=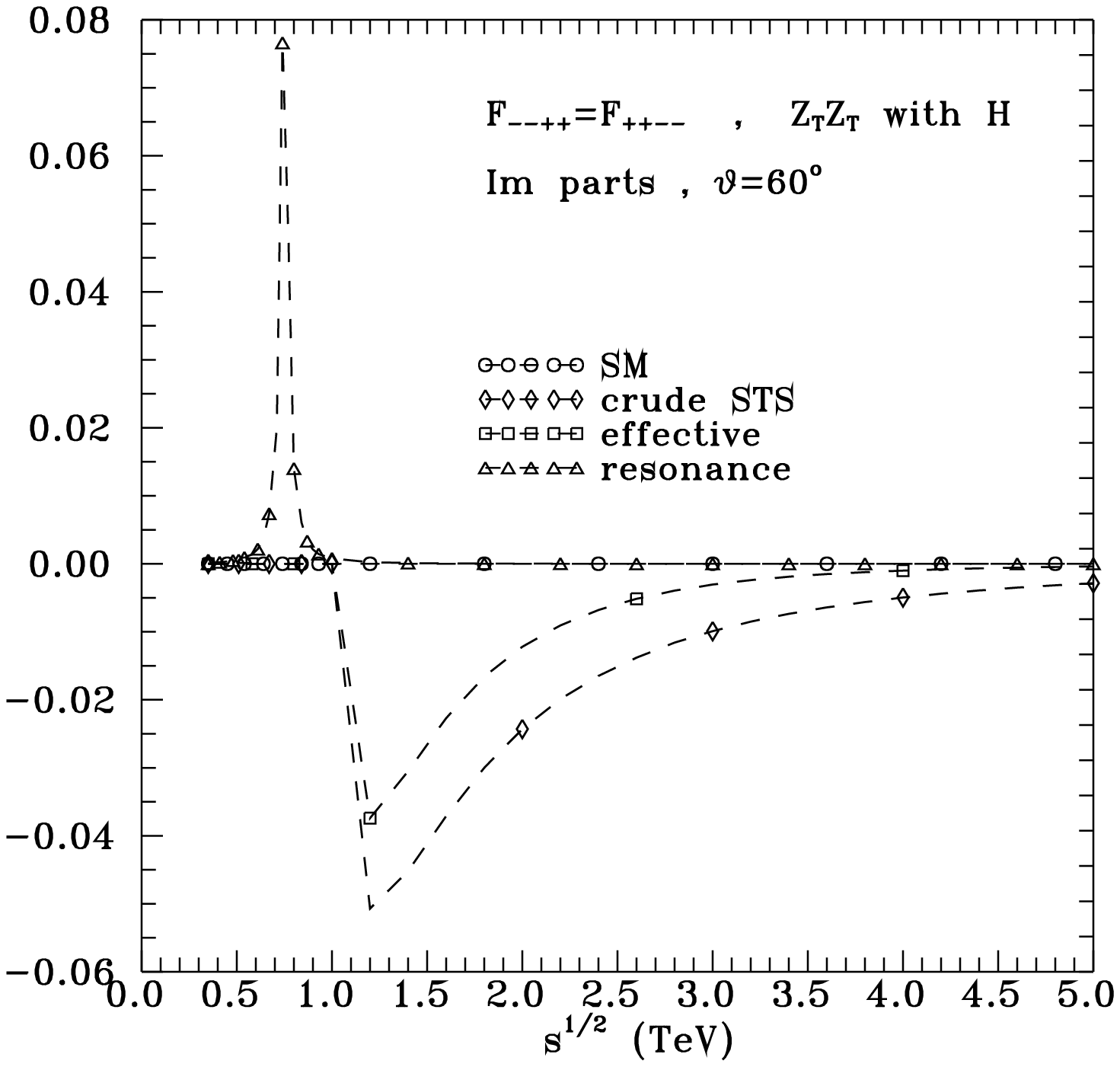, height=6.cm}
\]
\[
\epsfig{file=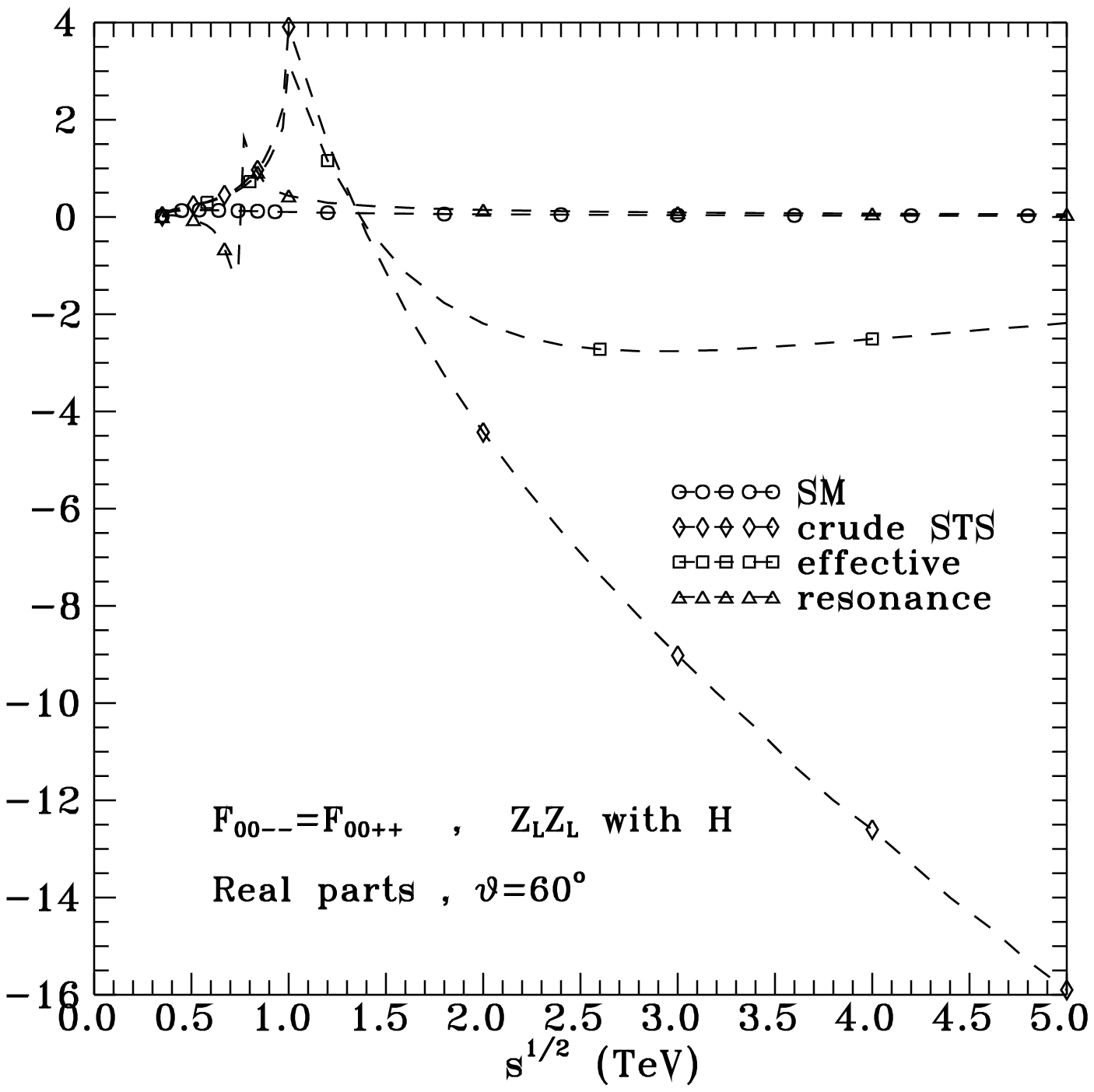, height=6.3cm}\hspace{0.5cm}
\epsfig{file=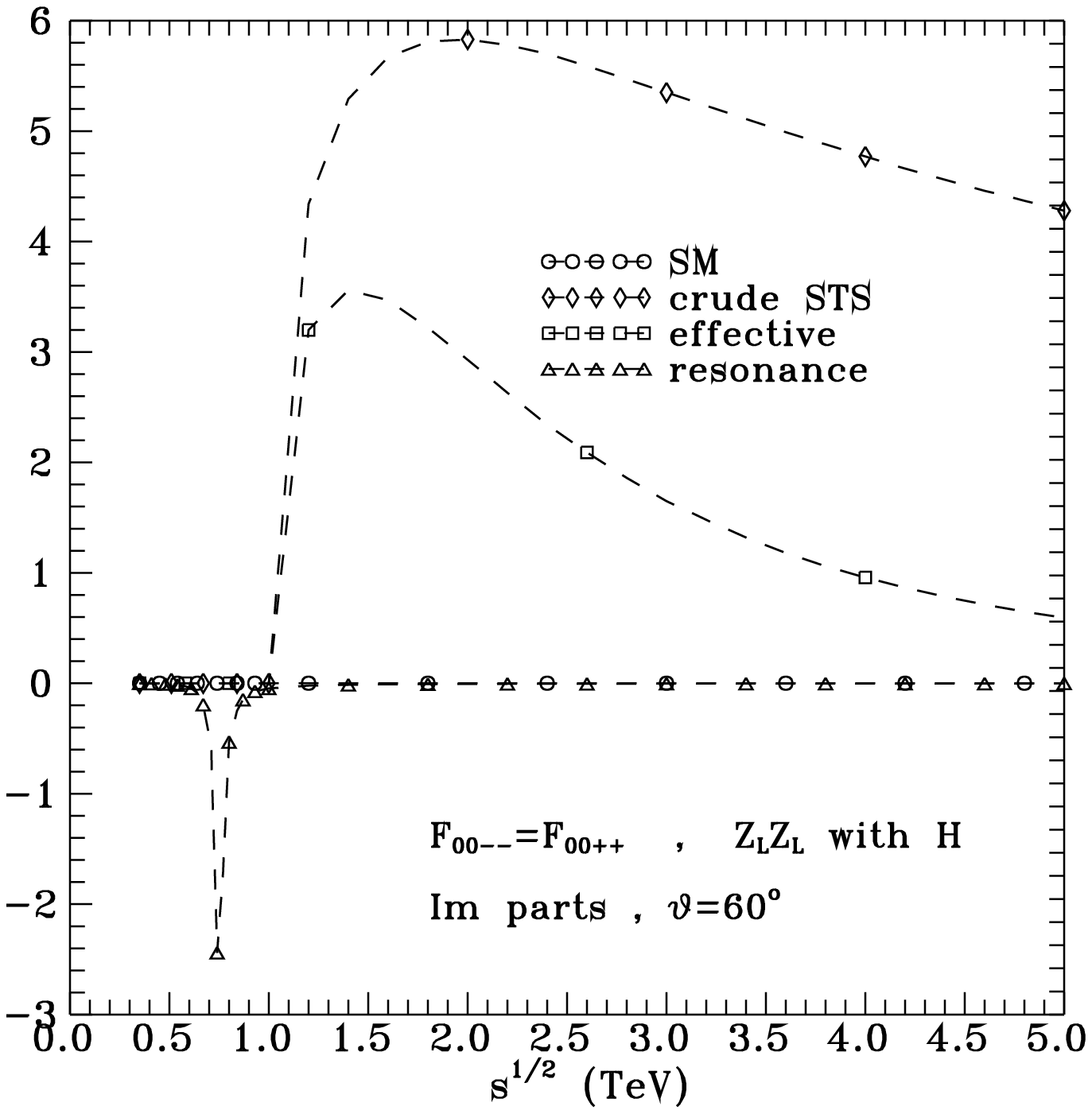, height=6.3cm}
\]
\caption[1]{ The SM results and the effects of the anomalous $Htt$ form factor
on the sensitive to it amplitudes $Z_\lambda Z_{\lambda'} \to t_\tau \bar t_{\tau'}$
 listed in (\ref{ZZ-HV2-SMamp}).
Real (imaginary) parts are shown in the left (right) panels respectively.
The definition of {\it crude STS, effective} and
{\it resonance} form factor models, as in Fig.1.}
\label{Fg6}
\end{figure}

\clearpage

\begin{figure}[p]
\[
\epsfig{file=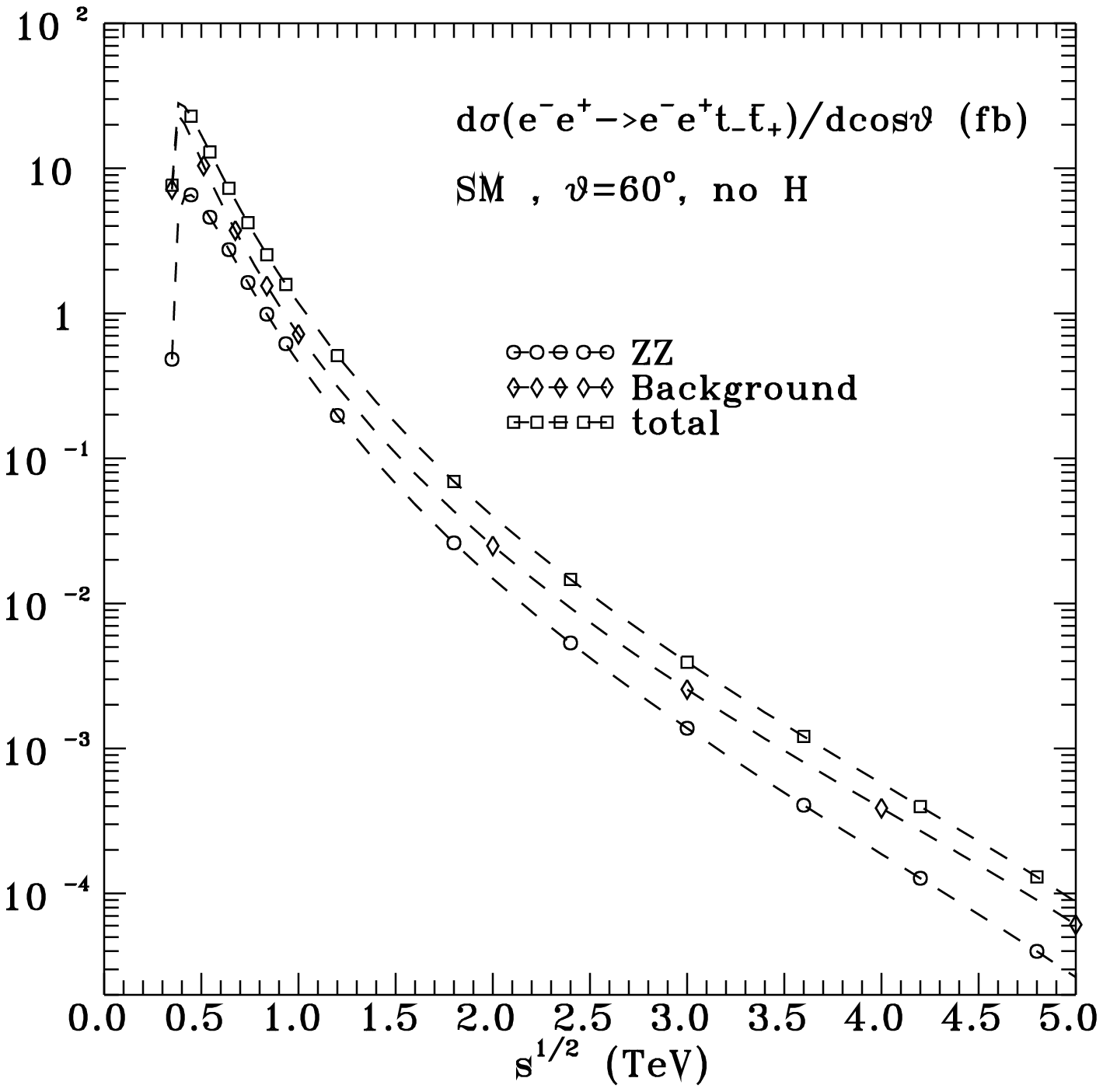, height=6.5cm}\hspace{0.5cm}
\epsfig{file=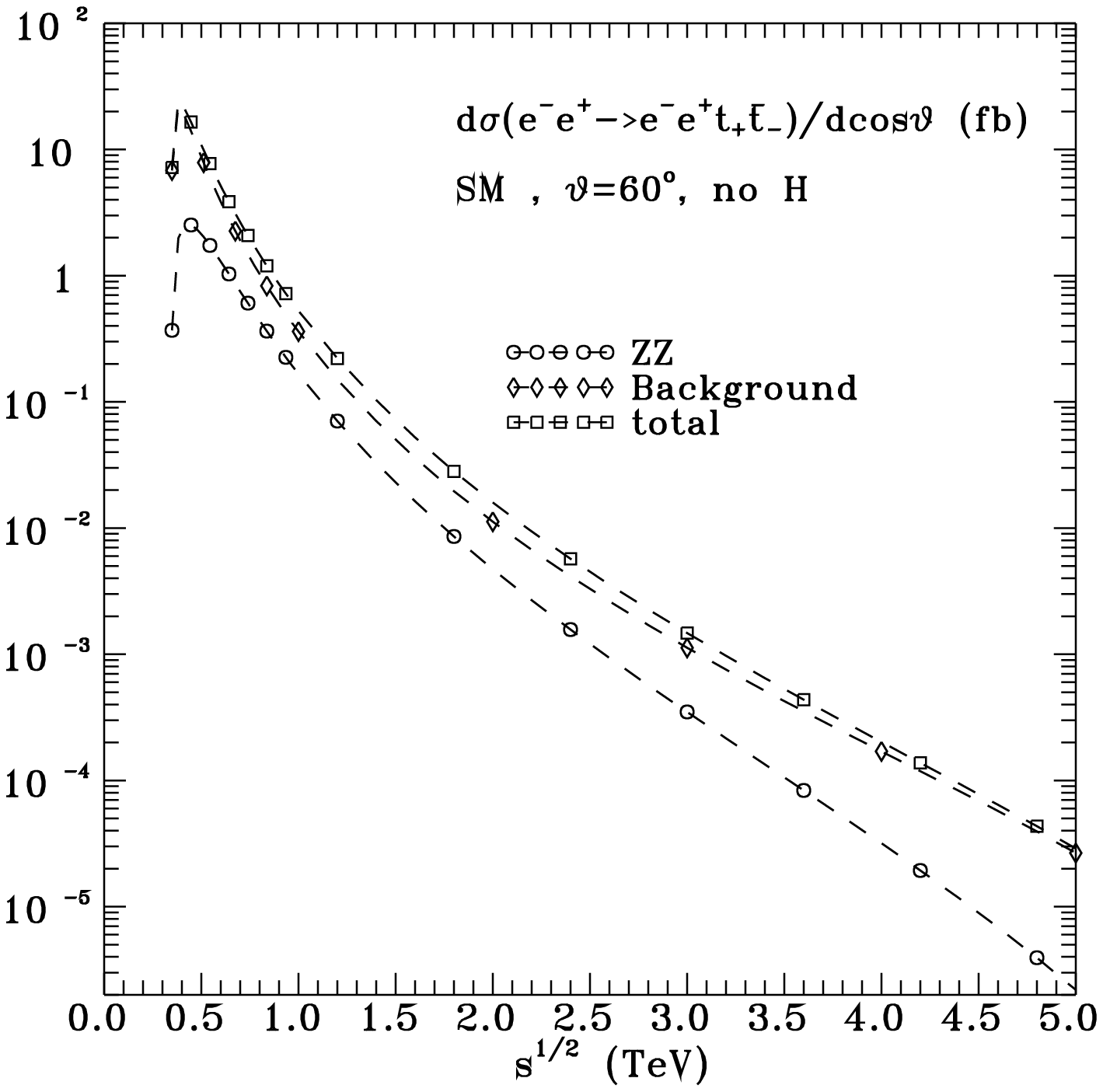, height=6.5cm}\hspace{0.5cm}
\]
\caption[1]{ Energy dependencies of the differential  cross sections for
$d \sigma (e^- e^+ \to e^- e^+  t_\tau \bar t_{\tau'})/d\cos\theta$ at  $\theta=60^\circ$.
The $ZZ$ intermediate state, the {\it background} induced by anything non-$ZZ$,
and the complete result denoted as "total" are separately given.
The left (right) panel corresponds to $\tau=-\tau'=- 1/2 ~(\tau=-\tau'=+ 1/2)$
and they both contain no $H$ contribution.}
\label{Fg7}
\end{figure}

\clearpage

\begin{figure}[p]
\[
\epsfig{file=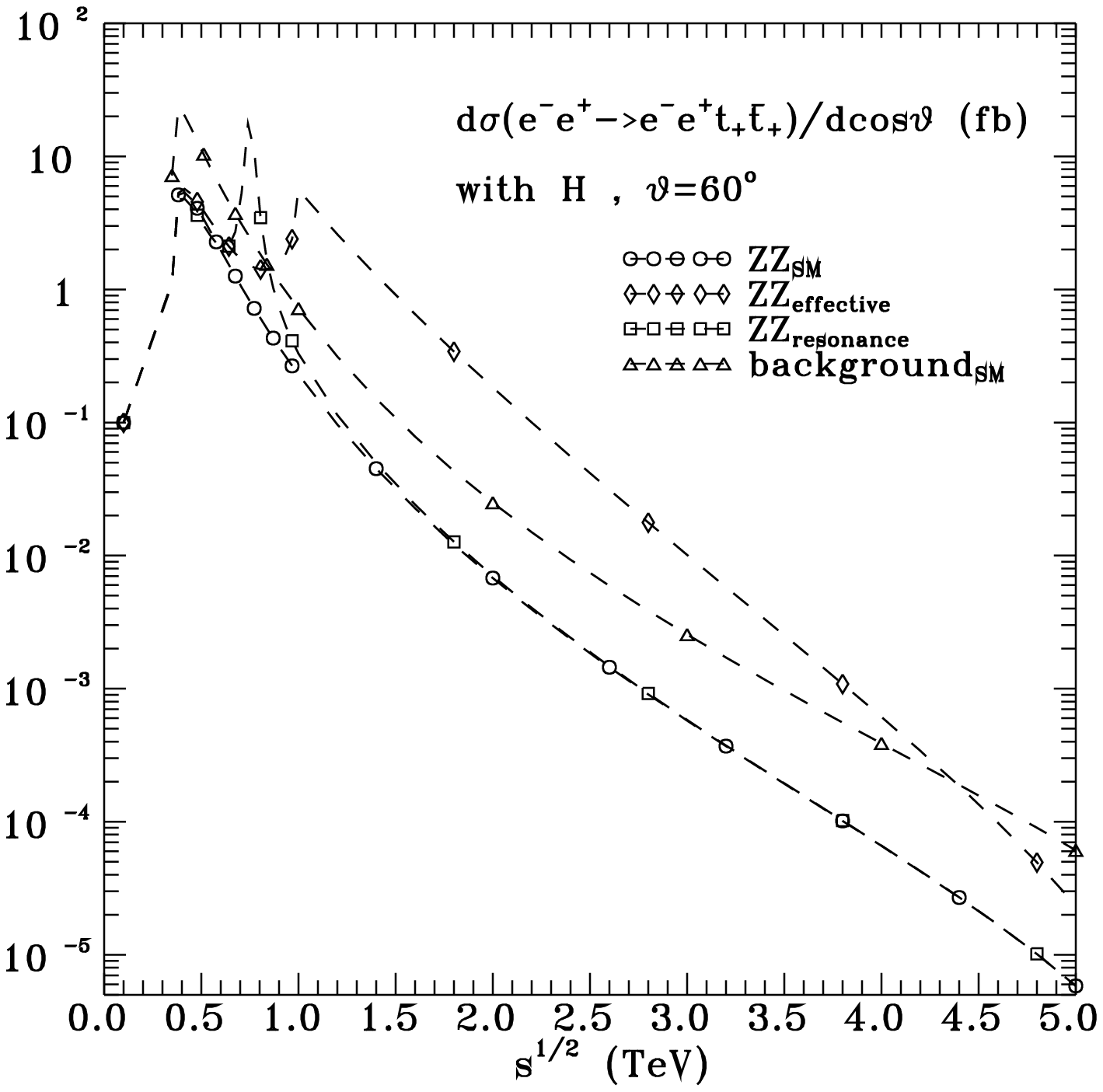, height=6.5cm}\hspace{0.5cm}
\epsfig{file=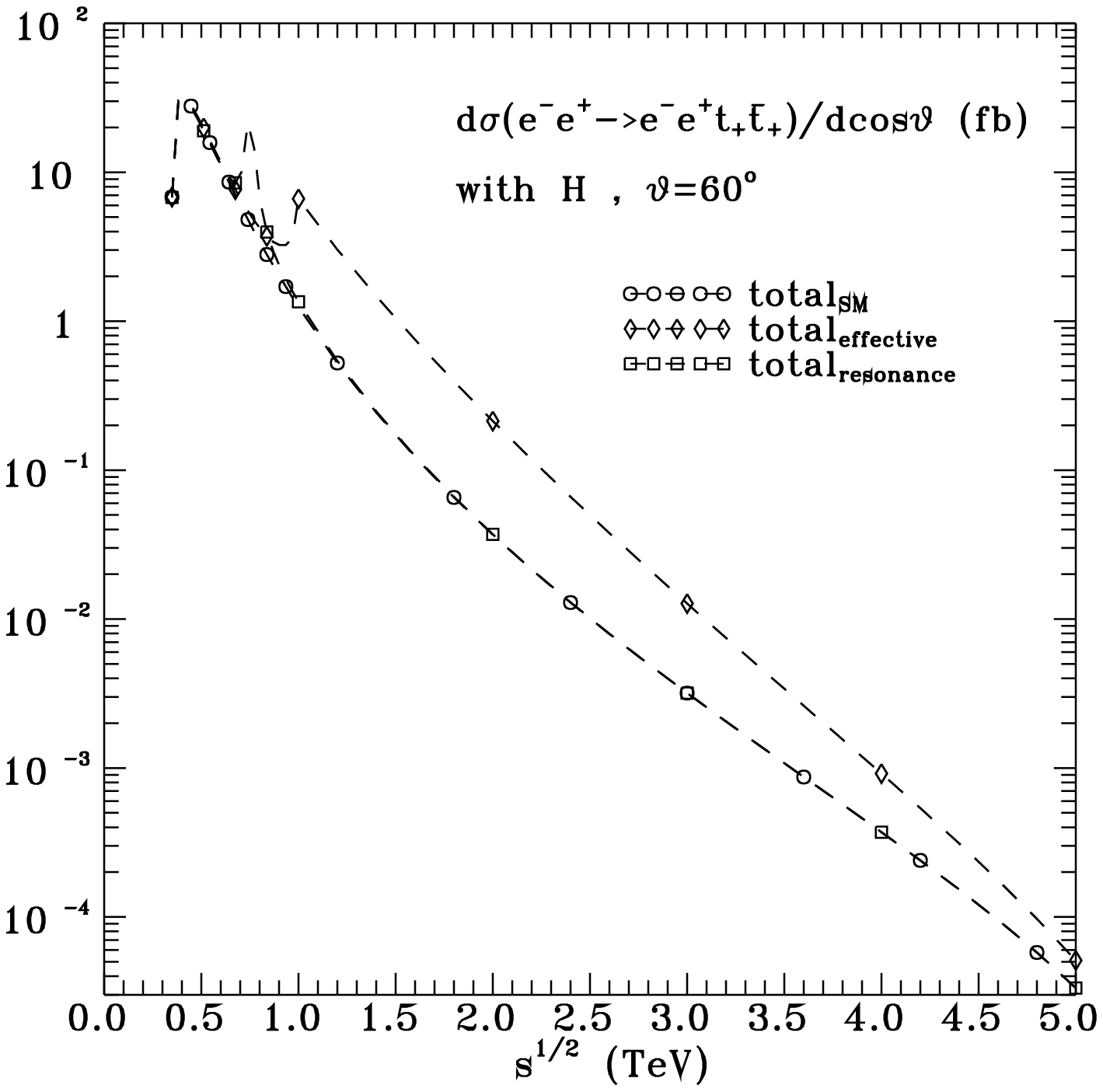,height=6.5cm}
\]
\[
\epsfig{file=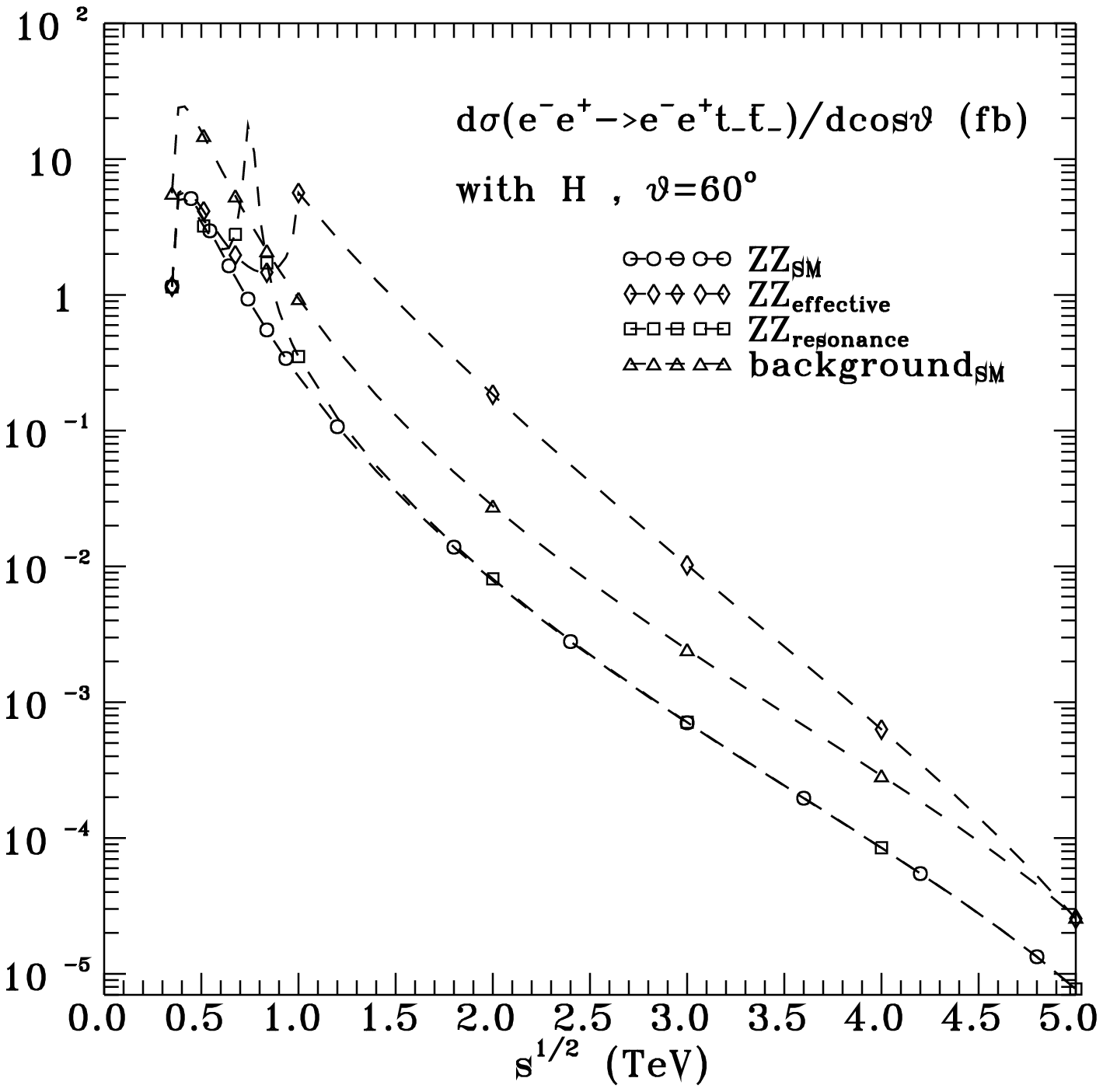, height=6.5cm}\hspace{0.5cm}
\epsfig{file=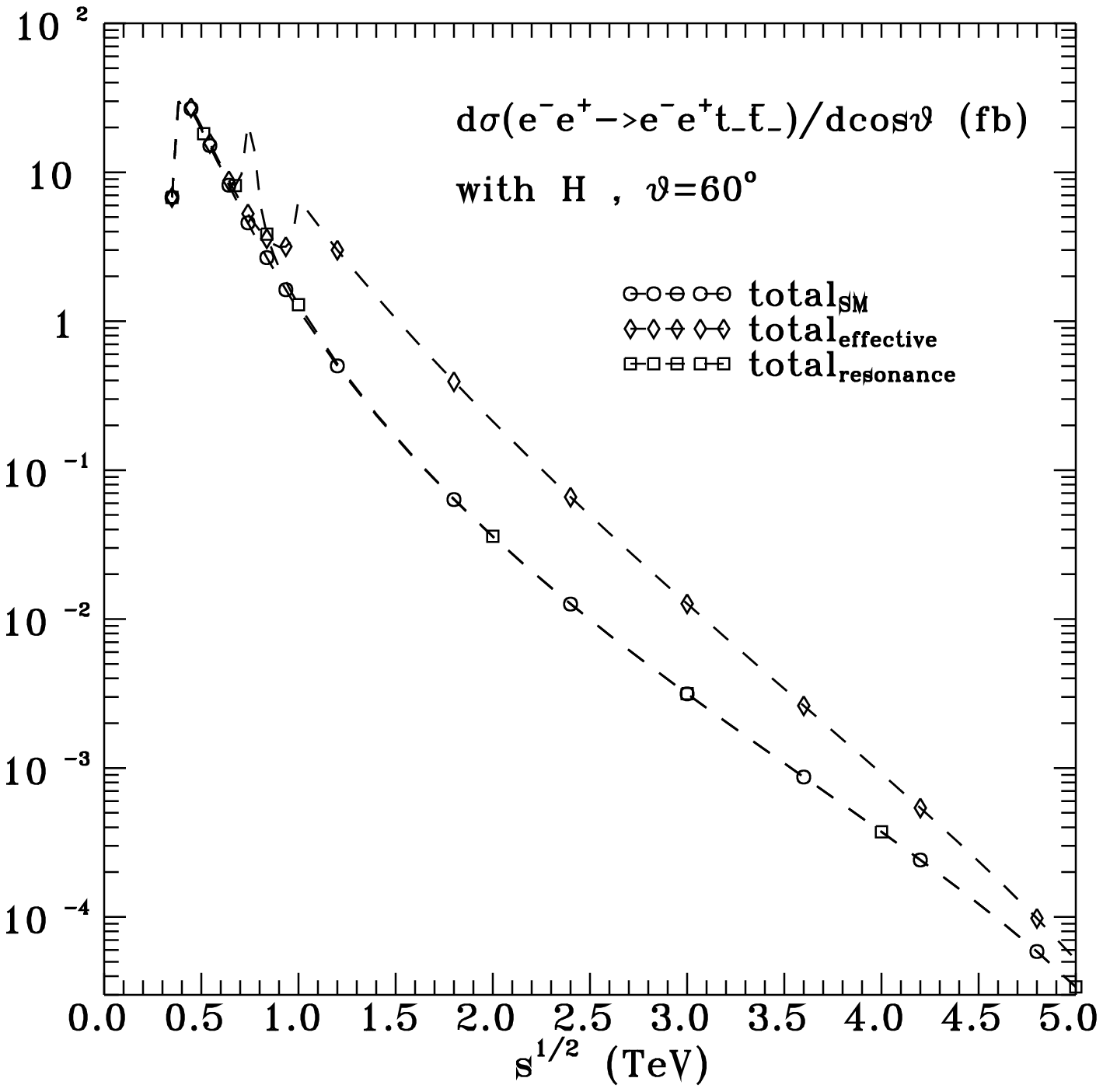, height=6.5cm}
\]
\caption[1]{Differential cross sections as in Fig.7, with
upper and lower panels respectively corresponding to $\tau=\tau'=\pm 1/2$
and both receiving  $H$ form factor contributions (see Fig.1) modifying the $ZZ$
(left)  and  the "total"  (right) panel results.}
\label{Fg8}
\end{figure}

\clearpage

\begin{figure}[p]
\[
\epsfig{file=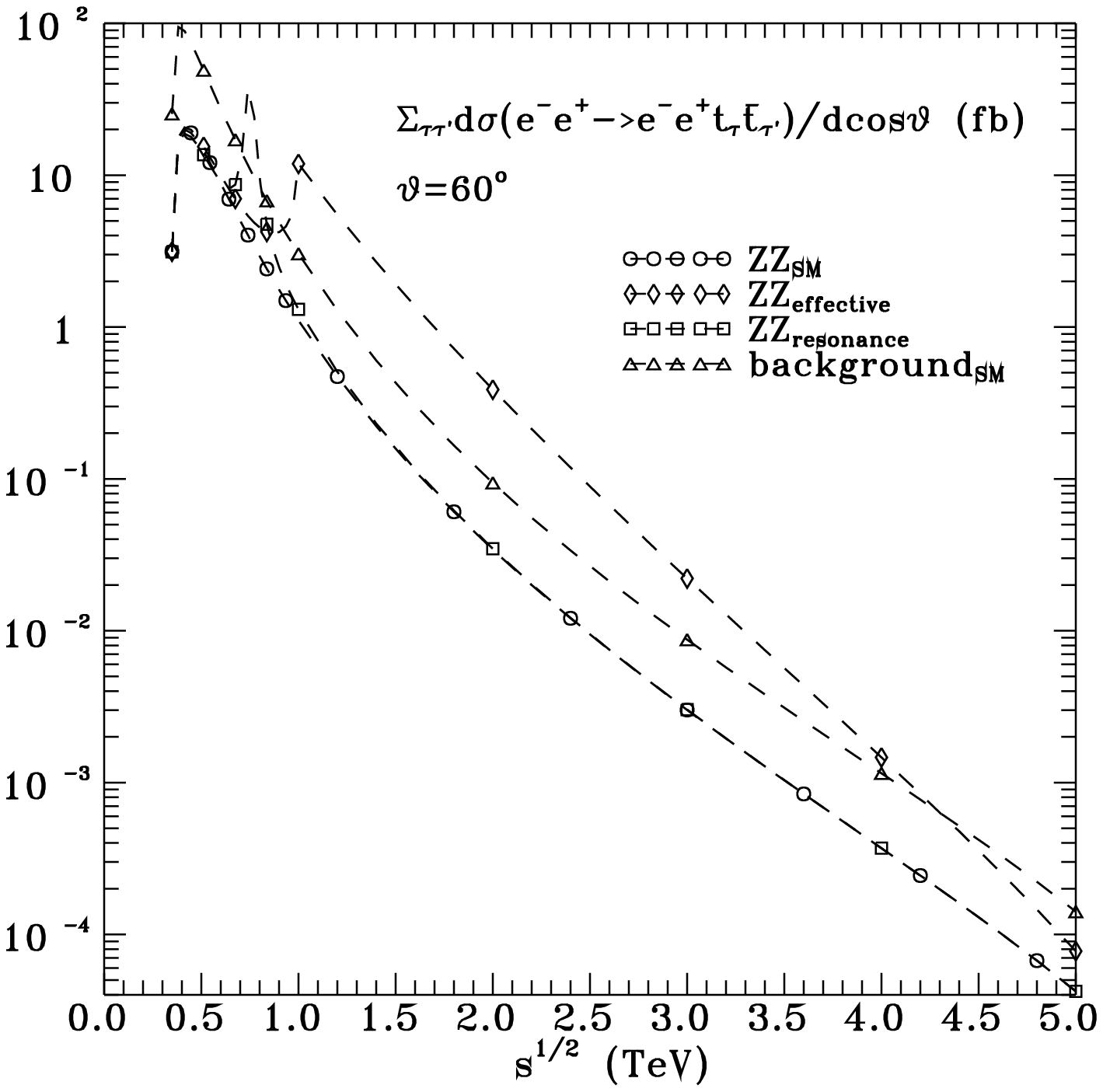, height=6.5cm}\hspace{0.5cm}
\epsfig{file=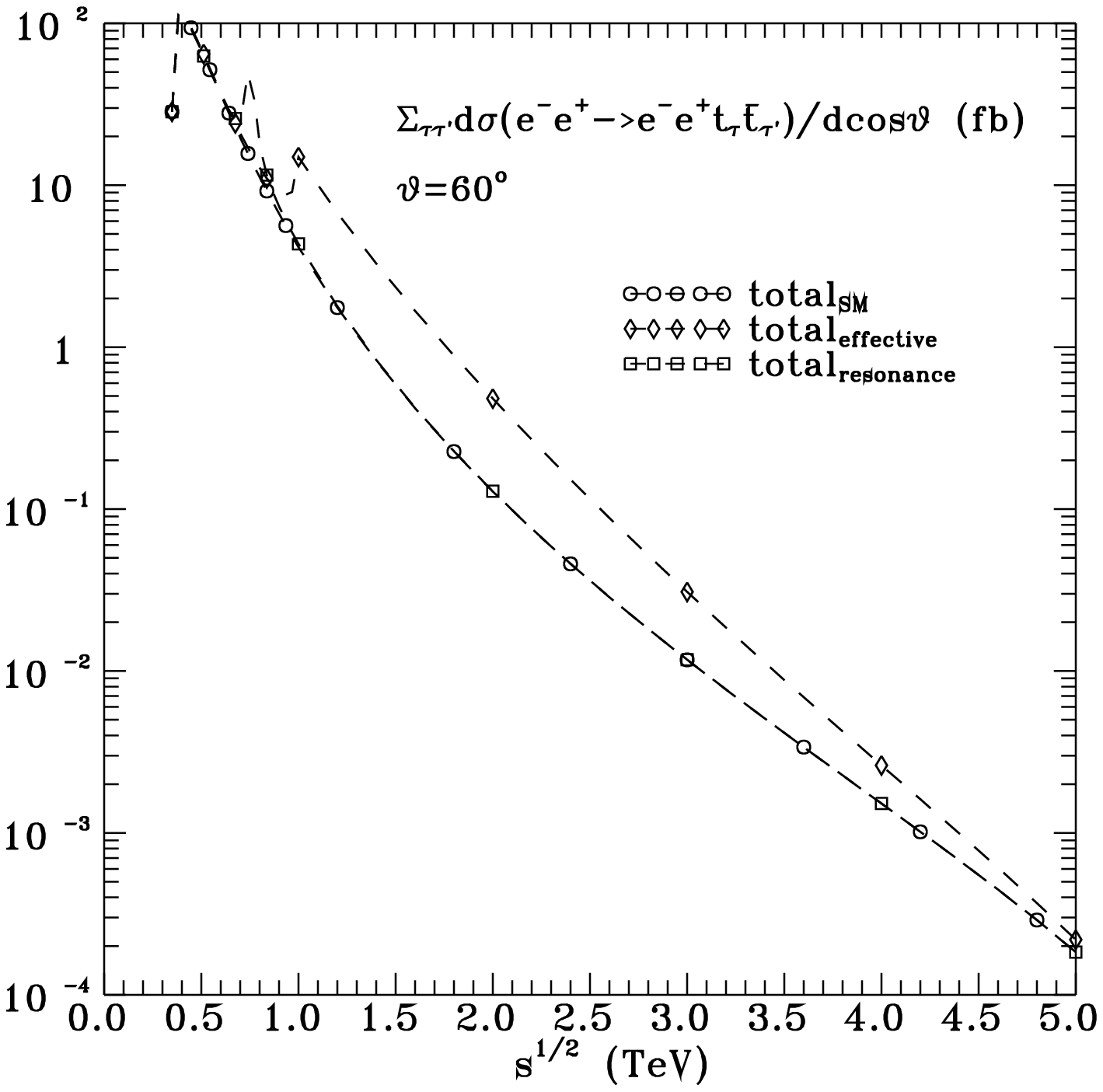,height=6.5cm}
\]
\[
\epsfig{file=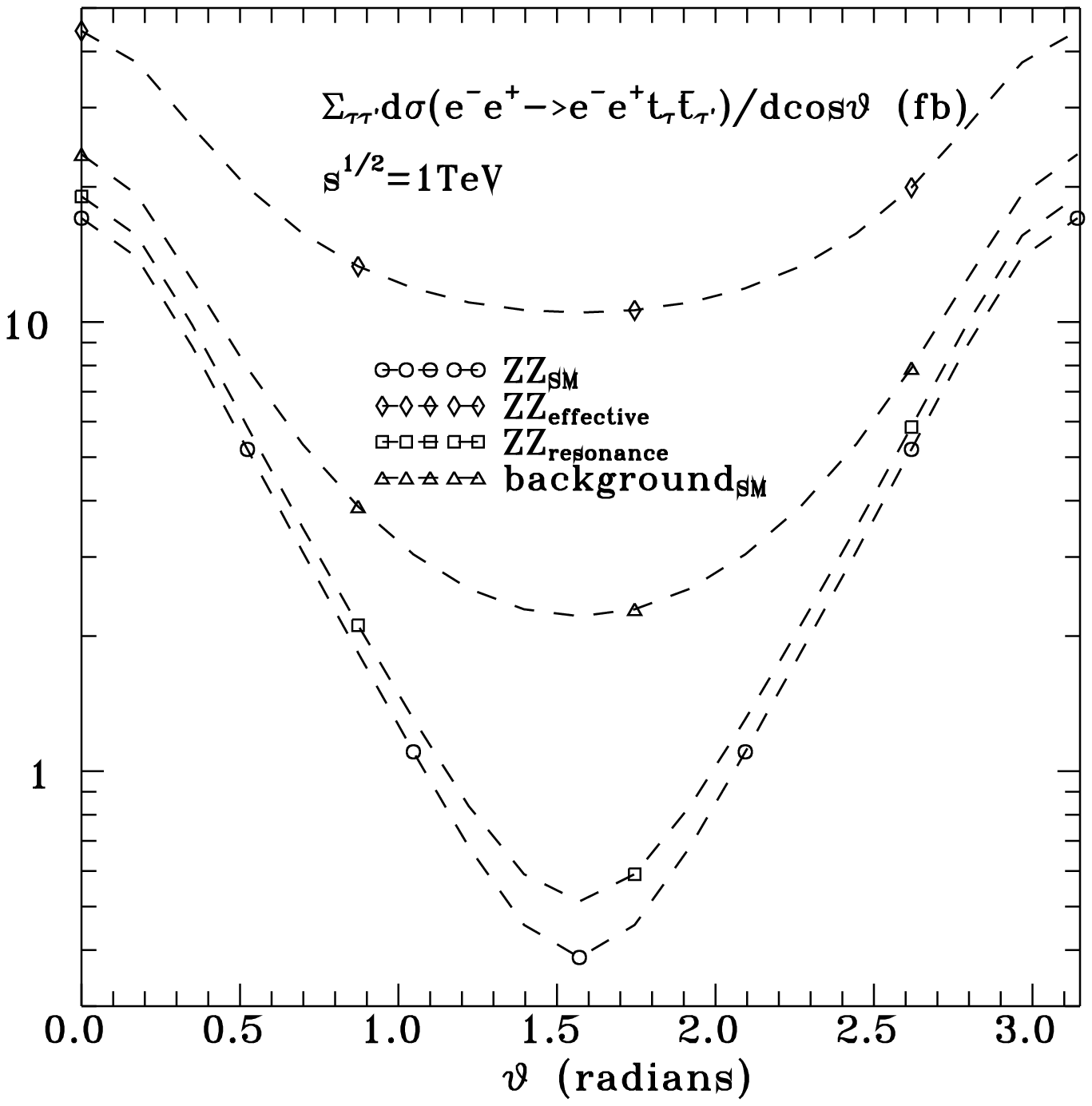, height=6.5cm}\hspace{0.5cm}
\epsfig{file=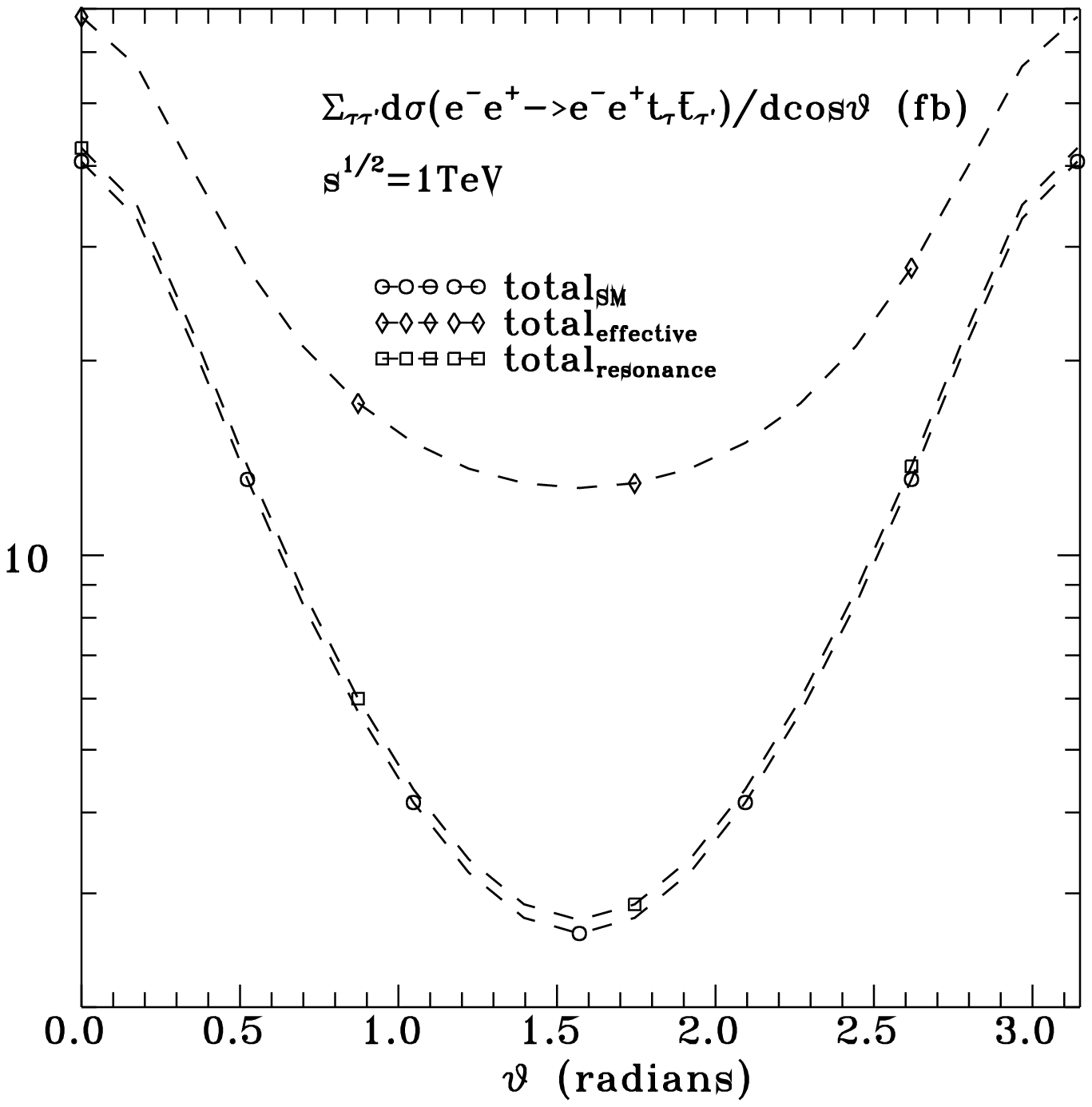,height=6.5cm}
\]
\caption[1]{Energy dependencies at  $\theta=60^\circ$ (upper panels) and angular dependencies
at $\sqrt{s}=1$TeV (lower panels)  of the unpolarized differential  cross sections for
$ \sum_{\tau \tau'}d \sigma (e^- e^+ \to e^- e^+  t_\tau \bar t_{\tau'})/d\cos\theta$.
The $ZZ$ intermediate state, the {\it background} induced by anything non-$ZZ$,
and the complete result indicated as "total" are separately given in SM and
after including the $H$ form factor (see Fig.1) contribution. }
\label{Fg9}
\end{figure}


\clearpage

\begin{thebibliography}{99}
%
\bibitem{BSMth}
M.E. Peskin, Ann.Phys.(N.Y.){\bf 528},20(2016). M. Muhlleitner, arXiv:1410.5093. Ben Gripaios,
arXiv:1503.02636, arXiv:1506.05039.
%
\bibitem{comp}  H. Terazawa, Y. Chikashige and K. Akama, \pr{D15}{480}{1977};
for other references see
H. Terazawa and M. Yasue, Nonlin.Phenom.Complex Syst. {\bf 19},1(2016);
\jmp{5}{205}{2014}.
%
\bibitem{Hcomp} G. Panico and A. Wulzer, Lect.Notes Phys. {\bf 913},1(2016).
%
\bibitem{Portal} B.Patt and F. Wilczek, arXiv: hep-ph/0605188.
%
\bibitem{Htop} R. Contino, T. Kramer, M. Son and R. Sundrum,
J. High Energy Phys. 05({\bf 2007})074.
%
\bibitem{fusion1} M.S. Chanowitz and M.K. Gaillard, \pl{B142}{85}{1984};
G.L. Kane, W.W. Repko and W.B. Rolnick,  \pl{B148}{367}{1984};
S. Dawson, \np{B249}{42}{1985}.
%
\bibitem{fusion2} R.P. Kauffman, \pr{D41}{3343}{1990}.
%
\bibitem{fusion3} M. Gintner and S. Godfrey, arXiv: 9612342 [hep-ph],
eConf.0960625(1996)STC 130, proceedings  of the
1996 Summer Study on New Directions for High-Energy Physics,  Snowmass 1996. C.-P. Yuan, \np{B310}{1}{1988}.
%
\bibitem{Djouadi} E. Accomando et al. \prep{299}{1}{1998}.
%
\bibitem{fusion4} M. Capdequi-Peyranere et al, \zp{C41}{1988}{99}.
%
\bibitem{JW} M. Jacob and G.C. Wick, \aop{7}{404}{1959}, \aop{281}{774}{2000}.
%
\bibitem{hc}  G.J. Gounaris and F.M. Renard,
\prl{94}{131601}{2005}; \pr{D73}{097301}{2006}.
%
\bibitem{Saavedra} J.A. Aguilar Saavedra, \np{B821}{215}{2009}.
%
\bibitem{PV} G. Passarino and M. Veltman, \np{B160}{151}{1979}.
%
\bibitem{SRS}  G.J. Gounaris and F.M. Renard,
 \polon{42}{2107}{2011};  \pr{D86}{013003}{2012}; \pr{D90}{073007}{2014}.
%
\bibitem{ATLAS750} G. Aad et al.[ATLAS Collaboration], Reports No. ATLAS-CONF-2015-081(2015) and No.
 ATLAS-CONF-2016-018.
%
\bibitem{CMS750} S. Chatrchyan et al.[CMS Collaboration], Reports No. CMS-PAS-EXO-15-004(2015) and No.
CMS-PAS-EXO-16-018.
%
\bibitem{no750} See for example
$http://indico.cern.ch/event/432527/contributions/1072336/attachments$
$/1321033/1981068/BL_ATLAS_HighMassDiphotons_ICHEP2016.pdf$\\
and
$http://indico.cern.ch/event/432527/contributions/1072431/attachments/1320985$
$/1980991/chiara_ichep.pdf$
%
\bibitem{Larios} F. Larios, T. Tait and C.-P. Yuan, \pr{D57}{3106}{1998}.
%
\bibitem{colliders} D. d'Enterria, arXiv: 1602.05043[hep-ex] and its Refs.1-4.
%

\end{thebibliography}
\end{document}